\documentstyle{mn}

\newif\ifAMStwofonts
\ifoldfss
  \ifCUPmtlplainloaded \else
    \NewTextAlphabet{textbfit} {cmbxti10} {}
    \NewTextAlphabet{textbfss} {cmssbx10} {}
    \NewMathAlphabet{mathbfit} {cmbxti10} {} % for math mode
    \NewMathAlphabet{mathbfss} {cmssbx10} {} %  "   "    "
  \fi
  \ifAMStwofonts
    \ifCUPmtlplainloaded \else
      \NewSymbolFont{upmath} {eurm10}
      \NewSymbolFont{AMSa} {msam10}
      \NewMathSymbol{\upi}     {0}{upmath}{19}
      \NewMathSymbol{\umu}     {0}{upmath}{16}
      \NewMathSymbol{\upartial}{0}{upmath}{40}
      \NewMathSymbol{\leqslant}{3}{AMSa}{36}
      \NewMathSymbol{\geqslant}{3}{AMSa}{3E}

      \let\leq=\leqslant 
      \let\geq=\geqslant 
    \fi
  \fi
\fi % End of OFSS

\ifnfssone
  \newmathalphabet{\mathit}
  \addtoversion{normal}{\mathit}{cmr}{m}{it}
  \addtoversion{bold}{\mathit}{cmr}{bx}{it}
  \newmathalphabet{\mathbfit} % math mode version of \textbfit{..}
  \addtoversion{normal}{\mathbfit}{cmr}{bx}{it}
  \addtoversion{bold}{\mathbfit}{cmr}{bx}{it}
  \newmathalphabet{\mathbfss} % math mode version of \textbfss{..}
  \addtoversion{normal}{\mathbfss}{cmss}{bx}{n}
  \addtoversion{bold}{\mathbfss}{cmss}{bx}{n}
  \ifAMStwofonts
    \ifCUPmtlplainloaded \else
      \UseAMStwoboldmath
      \makeatletter
      \new@mathgroup\upmath@group
      \define@mathgroup\mv@normal\upmath@group{eur}{m}{n}
      \define@mathgroup\mv@bold\upmath@group{eur}{b}{n}
      \edef\UPM{\hexnumber\upmath@group}
      \new@mathgroup\amsa@group
      \define@mathgroup\mv@normal\amsa@group{msa}{m}{n}
      \define@mathgroup\mv@bold\amsa@group{msa}{m}{n}
      \edef\AMSa{\hexnumber\amsa@group}
      \makeatother
      \mathchardef\upi="0\UPM19
      \mathchardef\umu="0\UPM16
      \mathchardef\upartial="0\UPM40
      \mathchardef\leqslant="3\AMSa36
      \mathchardef\geqslant="3\AMSa3E

      \let\leq=\leqslant 
      \let\geq=\geqslant 
    \fi
  \fi
\fi % End of NFSS release 1

\ifnfsstwo
  \DeclareMathAlphabet{\mathbfit}{OT1}{cmr}{bx}{it}
  \SetMathAlphabet\mathbfit{bold}{OT1}{cmr}{bx}{it}
  \DeclareMathAlphabet{\mathbfss}{OT1}{cmss}{bx}{n}
  \SetMathAlphabet\mathbfss{bold}{OT1}{cmss}{bx}{n}
  \ifAMStwofonts
    \ifCUPmtlplainloaded \else
      \DeclareSymbolFont{UPM}{U}{eur}{m}{n}
      \SetSymbolFont{UPM}{bold}{U}{eur}{b}{n}
      \DeclareSymbolFont{AMSa}{U}{msa}{m}{n}
      \DeclareMathSymbol{\upi}{0}{UPM}{"19}
      \DeclareMathSymbol{\umu}{0}{UPM}{"16}
      \DeclareMathSymbol{\upartial}{0}{UPM}{"40}
      \DeclareMathSymbol{\leqslant}{3}{AMSa}{"36}
      \DeclareMathSymbol{\geqslant}{3}{AMSa}{"3E}

      \let\leq=\leqslant 
      \let\geq=\geqslant 
    \fi
  \fi
\fi % End of NFSS release 2

\ifCUPmtlplainloaded \else
  \ifAMStwofonts \else % If no AMS fonts
    \def\upi{\pi}
    \def\umu{\mu}
    \def\upartial{\partial}
  \fi
\fi

\title{The CLASS blazar survey: testing the blazar sequence}
\author[A. Caccianiga \& M.J.M. March\~a]
       {A. Caccianiga$^1$ \& M.J.M. March\~a,$^2$  \\
$^1$INAF-Osservatorio Astronomico di Brera, via Brera 28, I-20121, Milano,
Italy\\
 $^2$CAAUL, Observat\'orio Astron\'omico de Lisboa, Tapada da Ajuda, 
            1349-018 Lisboa, Portugal 
}
\date{}

\pagerange{\pageref{firstpage}--\pageref{lastpage}}
\pubyear{1994}

\begin{document}

\input{psfig.txt}

\maketitle

\label{firstpage}

\begin{abstract} We discuss the properties of the sources in the CLASS
Blazar survey which aims at the selection of low radio power (P$_{5
GHz}<$ 10$^{25}$ W Hz$^{-1}$) blazars. We use VLA data from available
catalogues and from our own observations to constrain the radio
core-dominance of the sample which, together with the flat radio
spectral index, is a signature of the blazar activity. X-ray data from
the ROSAT All Sky Survey were also collected in order to constrain the
radio-to-X-ray luminosity ratio ($\alpha_{RX}$) of the sources. 
The data analysis shows that more than 30\% of sources at low radio power 
(P$_{5 GHz}<$10$^{25}$ W Hz$^{-1}$) have an
$\alpha_{RX}$ steeper than that expected in the framework of the
``blazar sequence'' recently put forward to unify the high and low power 
blazars. The
possibility that this result is influenced by contaminating sources in
the current sample is discussed. The conclusion is that, even if
a number of non-blazars (typically CSO/GPS sources) are expected in the
survey, it is unlikely that this constitutes the sole reason for the
observed deviation. In particular, we show 2 examples 
for which the blazar nature is confirmed from VLBI data and for
which the steep $\alpha_{RX}$ (suggesting a synchrotron peak frequency
below 10$^{15.5}$ Hz) and the low radio power (0.6-2$\times$10$^{24}$ 
W Hz$^{-1}$) put these sources outside the ``blazar sequence''.
The results presented here show the importance of a correct
and unbiased sampling of the low-power regime of the blazar population. 
%before deriving any conclusion about the existence of a correlation
%between the synchrotron peak frequency and the radio power.
%and suggest that
%a careful analysis of the content the ``blazar sequence'' in the
%low-power regime is the result of a selection effect due to the lack of
%blazar surveys sensitive to the low-power and steep $\alpha_{RX}$ sources.

\end{abstract}


\section{Introduction}

Blazars distinguish themselves from the rest of the radio-loud AGN by
their flat radio spectra, extreme variability, and high polarization
in both radio and optical wavelengths. These compact radio sources are
embedded in giant ellipticals galaxies and have their
Spectral Energy Distribution (SED) dominated by synchrotron and Inverse
Compton processes all the way from radio up to gamma-ray frequencies 
(see Urry \& Padovani 1995 for a review).
Blazars seem to come in two flavours: Flat Spectrum Radio
Quasars (FSRQ), which are more abundant at higher
luminosities, and BL Lacs which abound at lower
luminosities. The two types of sources differ
greatly in their optical spectroscopic properties. Whereas the former show
broad and strong emission lines, the latter are spectroscopically
dull sources, with only weak or absent emission lines.

Despite their spectroscopic differences, the SEDs of blazars have long
been modeled by two broad components: one synchrotron component
covering the frequencies from radio to optical, UV or even X-ray, and
another, ranging from X-ray to the $\gamma$ frequencies which is
attributed to Inverse Compton emission. It has been proposed
(Fossati et al. 1998; Ghisellini et al. 1998) that the blazar family
could be unified according to a single parameter related to the
bolometric luminosity. In particular, it was suggested that there is a
'blazar sequence' established by an anti-correlation between the
luminosity of the source and the frequency at which the synchrotron
component peaks. In this framework, following a decrease in luminosity
and an increase in frequency of the synchrotron peak we find, in
order, the FSRQs, the Low-energy peaked BL Lacs (LBL), and finally the
High-energy peaked BL Lacs (HBL). The astrophysical explanation put
forward by Ghisellini et al. (1998) for this continuity is that the
emitting particles suffer of increasing radiative losses as the
luminosity increases (see also Ghisellini, Celotti \& Costamante 2002). 
Thus, in the less intrinsically powerful sources
- the HBLs - the radiative cooling is less important and the highly
energetic particles can carry on producing synchrotron and Inverse Compton  
emission up to high frequencies. Conversely, the most powerful
sources - the FSRQs - suffer stronger cooling and synchrotron emission
peaks at much lower frequencies.

It must be noted,
however, that the blazar sequence rests on the observational lack of
low-power, low-energy peaked, and high-power, high-energy peaked
blazars. The question therefore arises: is the lack of such sources
real or is it a consequence of selection effects? In this latter
case, the seeming 'blazar sequence' could simply be due to limitations
of the blazar samples available at the time when the SEDs of the
different types of blazar were modeled by Fossati et al. (1998). 
If this is the case, deeper surveys able to
sample the 'radio power/peak frequency plane' more homogeneously
should find blazars outside the blazar sequence.

A good test to the proposed blazar continuity is to investigate its
validity in the low power regime, where primarily HBL should abound.
In this power regime, however, the usual classification of a BL Lac
or emission line object based on the optical spectral analysis is 
heavily hampered by the dilution of the nuclear emission due to
the presence of the host galaxy. This problem affects mostly the
LBL and this may result in a biased view of the blazar population at
these powers. An alternative tool to classify an object as blazar is thus 
required. In this paper we propose to use the radio compactness 
(on arcsecond scale) as an alternative way to select blazar candidates.
The aim of this paper is to present the analysis of the first large
and statistically significant sample of low-luminosity blazar 
candidates selected from the CLASS Blazar Survey (CBS, March\~a et al. 2001) 
on the basis of their radio compactness. This sample is used to 
test the validity of the blazar sequence for radio powers
below 10$^{25}$ W Hz$^{-1}$.

The paper is organised as follows: in the next Section 
we discuss the lack of samples able to select low-power blazars in an unbiased 
way. In the following two Sections (3 and 4) we briefly
present the selection criteria and optical data of the CBS.  In order
to avoid subtle bias, we will not make use of the optical data to
classify the sources as blazar. Instead, we exploit the radio data
(Section 5) to assess the blazar nature of the selected sources. The
X-ray data used to derive the X-ray-to-radio spectral indices are
presented in Section 6, and in Section 7 we discuss the consistency of
the data with blazar unified model.  In Sections 8, 9 and 10 we
investigate some explanations for the observed departure
from the blazar sequence. Finally, in Section 11 the summary
and conclusions are presented.  Throughout the paper the following
quantities are used: $H_{o}=$50~Km\,s$^{-1}$\,Mpc$^{-1}$ and
$q_{o}$=0.

\section{The need of a better sampling of the '$\alpha_{RX}$-P$_{5GHz}$ plane}

The blazar unified model (Fossati et al. 1998) has been 
proposed on the basis of the remarkable continuity observed in the 
SED behaviour as a function of the radio power. Three blazar samples 
were used to derive this correlation, namely the X-ray selected 
sample of 48 BL Lacs extracted from the {\it Einstein} Slew Survey
(Perlman et al. 1996), the radio selected sample of 34 BL Lacs
taken from the 1 Jy survey (Stickel et al. 1991) and the radio 
selected sample of 233 FSRQ of the 2 Jy survey (Padovani \& Urry 1992).

The key point of the blazar sequence rests on the lack of low-power 
(P$_{5 GHz}<$10$^{25}$ W Hz$^{-1}$) low-energy peaked ($\nu<$10$^{15.5}$ Hz) 
blazars and high power (P$_{5 GHz}>$10$^{26}$ W Hz$^{-1}$) 
high energy peaked blazars ($\nu>$10$^{16}$ Hz). 

In the low-power regime (which is the subject of the analysis presented 
in this paper) the lack of low-energy peaked
blazars is introduced by the lack of low-power blazars showing
a low X-ray-to-radio flux ratio (i.e. a ``steep'' radio-to-X-ray 
spectral index $\alpha_{RX}$\footnote{
The two point spectral index $\alpha_{RX}$ is here defined between 5~GHz
and 1 keV: 

$\alpha_{RX}=-Log(L_{5 GHz}/L_{1 keV})/Log(\nu_{5 GHz}/\nu_{1 keV})$

where $S_{5 GHz}$ and $S_{1 keV}$ are the monochromatic luminosities defined
at 5 GHz and 1 keV respectively and $\nu_{5 GHz}$ and $\nu_{1 keV}$ are
the corresponding frequencies. For the k-corrections and the 
conversion from the observed X-ray fluxes to the monochromatic 
fluxes we have assumed $\alpha_R$=0 and $\alpha_X$=1.
}): In fact, all the blazars considered by Fossati et al. (1998) with
a radio power below 10$^{25}$ W Hz$^{-1}$ have $\alpha_{RX}<$0.7
(see discussion in Section~8).

The important point we would like to stress here 
is that in the Fossati et al. (1998) 
analysis all the low-power blazars come from the Slew Survey 
which is strongly biased against steep $\alpha_{RX}$ objects,
due to the presence of an X-ray flux limit.

%the 1 Jy  and 2 Jy samples only contain high power objects 
%(between 10$^{25}$ W Hz$^{-1}$ and 10$^{30}$ W Hz$^{-1}$) due to the
%bright flux limits of these 2 samples. 
%The Slew Survey is sensitive mostly to HBLs, i.e. sources with
%a flat $\alpha_{RX}$.

To better clarify this point, let us consider a blazar with a radio power of 
10$^{24}$ W Hz$^{-1}$. Depending on the
$\alpha_{RX}$ value its X-ray luminosity (0.3-3.5 keV) ranges from
$\sim$6$\times$10$^{42}$ erg s$^{-1}$ ($\alpha_{RX}$=0.9) to
$\sim$6$\times$10$^{45}$ erg s$^{-1}$ ($\alpha_{RX}$=0.5).  Given the
flux limit of the Slew Survey ($\sim$5$\times$10$^{-12}$ erg s$^{-1}$
cm$^{-2}$) the detectability of an object with $\alpha_{RX}$=0.9 is
thus limited up to a low redshift, corresponding to a relatively
small volume of universe ($\sim$3$\times$10$^5$ Mpc$^3$, considering the area
of sky covered by the Slew Survey) when compared to the low BL Lac density
expected at these X-ray luminosities ($\sim$10$^{-6}$ Mpc$^{-3}$, based
on the extrapolation of the X-ray luminosity function presented in
Wolter et al. 1994). Thus, the Slew Survey is not expected to select such
steep low-power sources, even if they exist and have the same volume density as
the HBL objects.

%We conclude that, in an implicit way, the use of these 3 samples builds in 
%a radio power sequence, since from the start the used FSRQ were the most radio
%powerful sources, whilst the HBLs, the less radio powerful. 

A similar argument can be applied to the new samples based on the
cross correlation between radio and X-ray catalogues, like the 
the ROSAT Green-Bank Survey (RGB, Laurent-Muehleisen et al. 1998) and the
REX survey (Caccianiga et al. 1999). 
Even if these surveys go deep in the radio band ($\sim$5-50 mJy), the
requirement of an X-ray emission greater than few $\times$10$^{-14}$
erg cm$^{-2}$ s$^{-1}$ (REX), or even $\sim$10$^{-12}$ erg
cm$^{-2}$ s$^{-1}$ (RGB), makes these surveys not sensitive  to
steep $\alpha_{RX}$  low-power blazars.

The DXRBS is more sensitive to LBL-like objects, as
can be seen from the quite large number of sources with a steep
$\alpha_{RX}$ (e.g. see Padovani et al. 2003).  
However, this sample does not contain many 
low-power sources.
It is interesting to note that, even if based on small numbers, the DXRBS
shows a relatively large fraction ($\sim$33\%) of low-energy peaked
($\nu_{peak}<$10$^{15.5}$ Hz) BL Lacs in the low-power regime (P$_{5
GHz}<$2$\times$10$^{25}$ W Hz$^{-1}$) contrary
to what is expected from the blazar unified model (Padovani et al. 2003).

We conclude that the '$\alpha_{RX}-P_{5}$' plane has not been covered
homogeneously so far and, in particular, the low radio power regime
has been systematically sampled only for flat values of
$\alpha_{RX}$. Therefore, caution must be taken before 
deriving any conclusion about any power/peak frequency correlation.
The lack of any significant  anti-correlation between
the synchrotron peak frequency and the radio power among
the sources of the DXRBS recently found
by Padovani et al. (2003) seems to support this idea.

One of the goals of the CLASS Blazar Survey (CBS, March\~{a} et al. 
2001, herein paper~I) is to sample the low-power regime without a bias 
against steep $\alpha_{RX}$ blazars in order to verify the statements
about the blazars unified models.

\section{The CLASS blazar sample}

In March\~{a} et al. (2001, herein paper~I) the CLASS blazar sample (CBS)
has been presented and discussed. In summary, 325 flat spectrum radio
sources have been selected out of the CLASS survey 
(Myers et al. 2003) with the following criteria:

\begin{enumerate}
\item {$35\degr \leq\delta\leq75\degr$}
\item {$|b^{II}|\geq$20$^\circ$}
\item {$S_{5}\geq$ 30 mJy}
\item {flat spectrum, i.e. $\alpha_{1.4}^{4.8}\leq$0.5
(S$_{\nu}\propto\nu^{-\alpha}$)}
\item {red magnitude $\leq$ 17.5}
\end{enumerate}

A closer analysis of the radio maps have then excluded 23 ``fake''
flat-spectrum sources originally included in the sample as a result of
the different resolution of the radio catalogues (GB6 and NVSS) used
for the selection (see paper~I for the details). Thus, the final CBS
sample includes a total of 302 sources.

Even though the selection criteria summarized above are optimised to
select blazars, there remain some ``interlopers'' in the CBS. 
The exclusion of the 23 ``fake'' flat-spectrum sources described in
paper~I was based on the analysis of the NVSS data which does not
have the resolution required to exclude that some extended 
radio galaxies or quasars are still contaminating the sample. Also,
Compact Symmetric Objects (CSO), which have a compact morphology on
the kpc scales and often a flat radio spectrum (at the frequencies used for
the selection), can be ``confused'' with a blazar and hence included in the
CBS.  

In order to separate the ``real'' blazars in the sample from other
sources with physical properties unrelated to the blazar activity, we
have collected data over a wide range of different frequencies. In
principle, the most common signatures of the blazar activity, like the
variability, the optical polarization, or a featureless optical
continuum observed in a BL Lac object could be used to separate the
blazars from other contaminating sources in the CBS. However, at the
radio power regimes considered here ( between 10$^{23}$ and 10$^{25}$
W Hz$^{-1}$), the optical emission of a blazar is expected to be
easily overwhelmed by the light of the host galaxy in a significant
fraction of cases. Thus, direct evidence of the presence of a
featureless synchrotron continuum is difficult to obtain. Even if
these problems are not expected to affect all the selected blazars, it
is important to note that the blind use of the optical
features to classify an object as blazar could introduce subtle
selection effects. For this reason, we make use primarily of the radio
properties to constrain the presence of a blazar in the CBS
sources. The optical observations are used to determine the redshift
of the sources.  For completeness, the phenomenological classification 
based on the optical data is also briefly summarized in the next
section.

\section{The optical data}

So far, optical classification has been collected, either from the
literature or from specific observations for 91\% of the objects in
the sample.  The majority of objects with spectroscopic
classifications are presented in Caccianiga et al. (2001, hereafter
paper~II) together with a discussion of the criteria used to classify
the sources. Basically, the objects have been broadly divided into
those with no emission lines (type~0 objects), including BL Lacs and
Passive Elliptical Galaxies (PEGs), objects with broad emission lines
(type~1), and objects showing only narrow emission lines
(type~2). Since paper~II was published, however, new observations have
been carried out. The current percentage of sources in the three
classes is, respectively, 44\% (type~0), 40\% (type~1) and 16\%
(type~2).

As we have already mentioned, the present optical
classification is very sensitive to the contribution of the
host-galaxy in the observed spectrum. Given the nature of the sample,
which is mainly focused on the selection of low-power and low redshift 
sources, the host galaxy can even overshine the nuclear emission
making the spectroscopical classification of the source very
difficult. This problem affects mainly the featureless objects (BL Lac
objects) but also weak emission lines can be undetected because of the
host galaxy. 

The problem of emission line strength in the classification of
low-power blazars has no simple solution and a specific discussion
will be presented in a forthcoming paper. In the present work, instead,
we concentrate on the fact that the essential common feature of all
blazars is their flat radio spectrum and radio compactness. 

\section{radio data: Establishing the compactness of the sample}

Blazars are thought to be radio galaxies whose jet is well aligned with
the line of sight, and therefore highly affected by relativistic
beaming (e.g. Urry \& Padovani 1995). 
It is in this framework that their flat radio spectra are
interpreted as the superposition of several compact, self-absorbed
components in the jet. Hence, blazars should have their radio
emission dominated by a compact region. In this Section we
analyse the compactness of sources in the CBS using VLA data
taken in different configurations and frequencies.

\subsection{The VLA B-array data}

The GB6 and NVSS data available by definition for all the sources in
the sample are not very effective  to distinguish the compactness of the
sources. For instance, the large beam of the NVSS (FWHM=45$\arcsec$)
corresponds to a linear scale of $\sim$300 kpc at z=0.3.

Observations in the B configuration of VLA represent the best
compromise in terms of resolution and sensitivity to extended
structures to allow the study of the radio ``compactness'' of the CBS
sources.  The resolution in this configuration ($\sim$5\arcsec at
1.4~GHz) corresponds to a linear scale of about 30 kpc (z=0.3) and
thus allows us to determine whether the radio emission comes from
within the host galaxy or from some more extended structure. At this
resolution a typical radio galaxy will appear clearly extended. On the
contrary, a core-dominated source will be unresolved. At the same
time, the largest angular scale ($\theta_{LAS}$) detectable in this
configuration and frequency (about 2$\arcmin$) is large enough to
avoid any flux loss for the majority of the CLASS sources which are
typically unresolved (integrated flux $\sim$ peak flux) at the NVSS
resolution. For the few sources clearly extended even at the NVSS
resolution we can use the NVSS integrated flux which is reasonably
representative of the total flux ($\theta_{LAS}$=15$\arcmin$).

For 61\% of the entire CBS sources data
from the FIRST survey (Becker et al. 1995) are available. 
The FIRST survey (Faint Image of Radio Sky at Twenty centimeters)
is made with the B-array of VLA at 1.4~GHz reaching an rms. of
$\sim$0.15 mJy/beam. In particular, we have used the most recent version 
of the FIRST catalogue (April 2003). We have collected additional VLA
data with the same configuration and frequency, and similar rms as
FIRST. Observations were made on May 11th and 27th, 2001 (24 hours
in total) of 68 sources (48 with z$<$0.3).  The typical
time-on-source of these snapshots ranges from 4 to 5 minutes reaching
an rms of 0.1-0.6 mJy/beam. The calibrators used during the
observations of May 11th were J0029+349 and J0137+331 while the
calibrators used in May 27th were J0614+607, J0921+622, J1331+305,
J1435+760 and J1927+612. The data have been reduced using the AIPS
(Astronomical Imaging Processing System) package of the National Radio
Astronomy Observatory (NRAO).

In Table~1 the VLA B-array data are presented. In total, VLA B-array
1.4~GHz data are now available for 78\% of the entire CBS sample, and
for 84\% of the sources with z$<$0.3.  Apart from 7 objects that
turned out to have a double morphology (see Table~1), all the
remaining objects show a clear core plus (sometimes) one sided
extended emission. Using these data we have computed a core-dominance
parameter (R) defined as follows:

\begin{center}
R = $\frac{S_{core}}{S_{ext}}\times$(1+z)$^{-1}$
\end{center}

where we have assumed the core flux density (S$_{core}$) equal to the
peak flux density of the brightest component, and the extended flux
density ($S_{ext}$) equal to the total flux density minus the peak
flux density.  The total flux density is the integrated flux density
computed by us or derived from the FIRST catalogue. If more than one
component is present in the FIRST catalogue the total flux density is
the sum of all these components. For the few extended sources 
(46 in total, 37 with z$<$0.3) where a fraction of the flux
density could have been missed in the VLA B-configuration, we have
used the NVSS integrated flux density as total flux density.  
These objects are clearly among the less core-dominated ones: 
the large majority (96\%) of these sources have R$<$10 and the
average value of R is equal to 1.6.

Finally, for the 7 sources with a double morphology, we have
considered the 8.4~GHz flux taken at the VLA in A-array (see next section) 
as core flux if the source is detected 
at this frequency. If the source is not detected at 8.4~GHz,
we have considered an upper-limit on the core flux equal to 1.4~mJy.

The factor (1+z)$^{-1}$ assumes a difference in spectral index between
core and extended emission of unity and takes into account, in first
approximation, the different spectral indices of the core and extended
components (Urry \& Padovani 1995).  Note that all the quantities are
computed at 1.4~GHz given the availability of the data at this
frequency.

In seven cases the core flux is equal to the total flux or slightly
larger, due to statistical fluctuations in the measuring algorithm. In
these cases a lower limit on the core dominance parameter has been set
by assuming an error on the computed fluxes of 0.1 mJy which
corresponds to a lower limit on R equal to core flux/0.1 mJy.

\begin{figure} 
\centerline{
\psfig{file=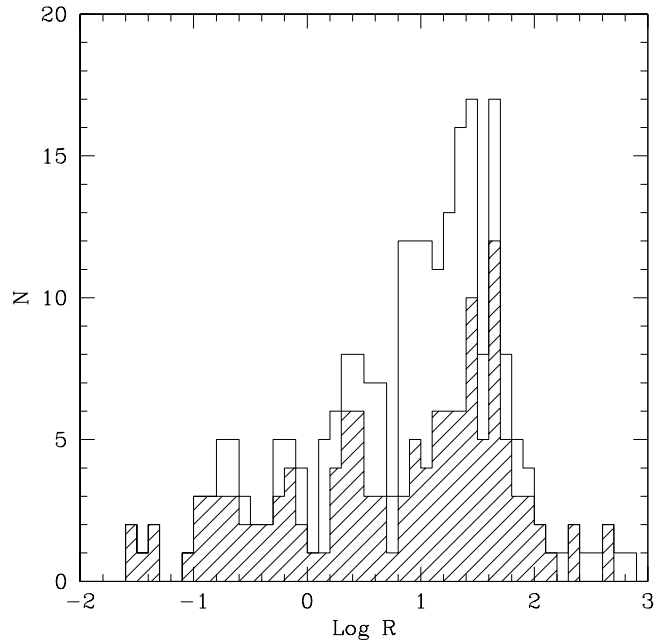,height=9cm,width=9cm} 
} 
\caption{The distribution of the core-dominance parameter
computed as described in the text for all the sources for which 
VLA B-array data at 1.4~GHz are available. The shaded area indicates
the sources with z$<$0.3} 
\label{r_dist} 
\end{figure}

\subsection{The VLA A-array data}

As described in paper~I, all the CLASS sources have been
systematically mapped at 8.4~GHz with the VLA in A-configuration (see
also Myers et al. 2003 for a detailed description of the data).  All
the 8.4~GHz fluxes of the CBS sources have been reported in paper~I.
The resolution achieved with these observations is about 0.24
arcsecond, hence a factor 20 higher resolution than the B-array
data. These data are useful to further constrain the compactness of
the sources although it must be kept in mind that flux on scales
larger than 7~arcsec can be lost in this configuration. Thus a steep
spectral index between 1.4 and 8.4~GHz is an indication that the
source is not very compact on $\sim$arcsecond scale.

\begin{figure} 
\centerline{ 
\psfig{file=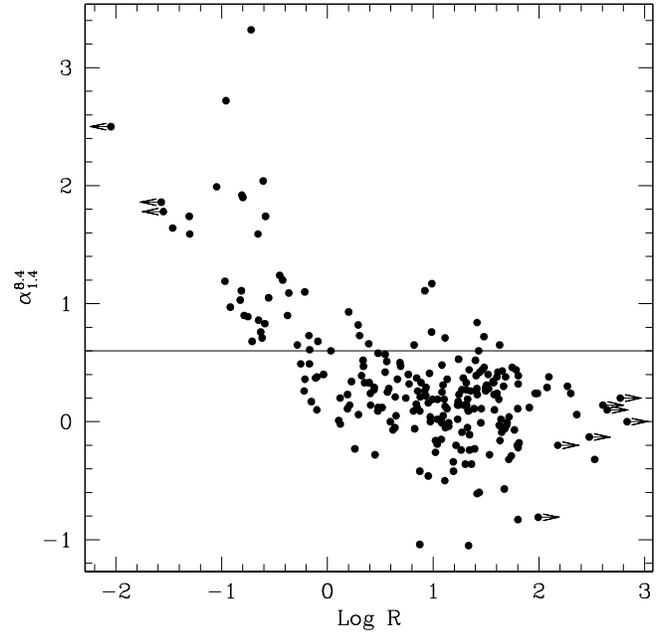,height=9cm,width=9cm}
} 
\caption{The radio spectral index 
between 1.4~GHz and 8.4~GHz versus the core-dominance parameter}
\label{r_a18} 
\end{figure}

In Figure~\ref{r_a18} the core-dominance parameter computed for the
sources with VLA B-array data is plotted against the spectral index
($\alpha_{1.4}^{8.4}$) between 1.4~GHz (using the NVSS total flux) and
8.4~GHz.  The evident correlation between the two quantities supports
the idea that the $\alpha_{1.4}^{8.4}$ can be used to constrain the
compactness of the sources, when the VLA B-array is not available. 
%All
%but two sources which are core dominated (R$>$1) have a spectral index
%flatter than 1, while all the very steep sources
%($\alpha_{1.4}^{8.4}>$1) have R$<$1.
In particular, based on Figure~\ref{r_a18} we find that 
the large majority (95\%) of the sources with 
$\alpha_{1.4}^{8.4}<$0.6 (continuous line in Figure~\ref{r_a18}) 
have R$>$1. At the same time, the large majority (94\%) of the
sources with R$>$1 have $\alpha_{1.4}^{8.4}<$0.6.
Thus, for the sources for which the VLA B-array is not available
the constraint $\alpha_{1.4}^{8.4}<$0.6 is almost equivalent to require R$>$1. 

\begin{figure} 
\centerline{ 
\psfig{file=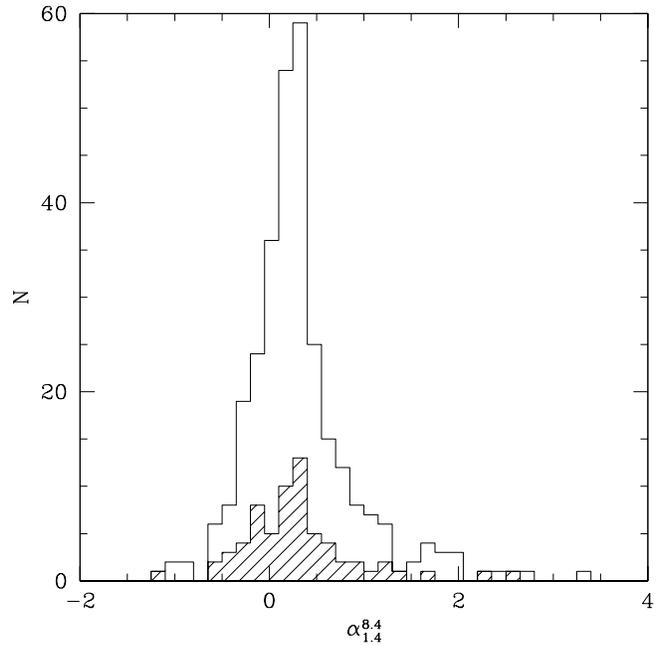,height=9cm,width=9cm}
} 
\caption{The distribution of the radio spectral index 
between 1.4~GHz and 8.4~GHz for all the sources in the CBS and
for those with no VLA B-array data available (shaded histogram)}
\label{a18} 
\end{figure}

In Figure~\ref{a18} the distribution of $\alpha_{1.4}^{8.4}$ is 
presented for all the sources in the CBS and for those with 
no VLA B-array data available (shaded histogram). Only about 20\% of
the sources without VLA B-array data have $\alpha_{1.4}^{8.4}>0.6$ 
and are thus expected to have (mostly) R$<$1.

In conclusion, the analysis of the VLA data collected so far
strongly indicates that the majority ($\sim$80\%) of the sources selected
in the CBS sample is actually core-dominated (R$>$1).

\subsection{A working definition of blazar candidate}

As already described, our aim is to use the radio data to
pin-point the  blazars selected in the CBS. By using
the VLA data described above, we consider an object as ``bona-fide'' 
blazar if R$>$1 or, in case the B-array data are not present, if
$\alpha_{1.4}^{8.4}\leq$0.6.

In the context of the beaming model,
the R parameter is directly linked to the beaming parameter ($\delta^p$
\footnote{ 
$\delta$ is the Doppler factor defined as $[\Gamma(1-\beta cos\theta)]^{-1}$,
where $\Gamma$ and $\beta$ are the Lorentz factor and the ratio between
the bulk velocity and the light speed, respectively. 
The exponent p is equal to $\alpha$+2 for a continuous jet 
where $\alpha$ is the spectral index (see Urry \& Padovani 1995 for 
a detailed description of the beaming model).
}):
\begin{center}
$R = f \delta^p$
\end{center}

where f is the ratio of intrinsic jet to unbeamed luminosity 
(Urry \& Padovani 1995). Assuming the fiducial values for f(=0.01), 
$\Gamma$ (=5) and by using the mean $\alpha$ for the sample (=0.4), 
the imposition of R$>$1 corresponds to observing angles below $\sim$8 degrees.
Hence, the definition
of a blazar based on the R parameter has a direct physical
meaning in terms of degree of beaming  and viewing angle.

Based on this definition, 244 sources in the CBS sample are thus
classified as blazar candidates.  In the following sections the term blazar
candidate will
be used to specify these 244 sources whose properties will be analysed
and discussed.

\section{The X-ray data}

In order to collect X-ray information for the CBS sources we have used the 
publicly available ROSAT All Sky Survey (RASS) catalogue 
which reaches flux levels of about 10$^{-12}$
erg\,s$^{-1}$\,cm$^{-2}$ in the 0.1-2.4 keV energy band (the Bright
Source Catalogue, BSC, Voges et al. 1999), and $\sim$5$\times$10$^{-13}$
erg\,s$^{-1}$ its faint extension (the Faint Source Catalogue, FSC,
Voges et al., http://www.xray.mpe.mpg.de/rosat/survey/rass-fsc/).

We have thus cross-correlated the CBS against the two RASS catalogues
with a positional tolerance of 40$\arcsec$. The reliability of the
cross-correlation has been assessed by performing the same 
cross-correlation after having positionally shifted the two catalogues
one from each other by an offset much larger than 40$\arcsec$, 
so that only chance coincidences are obtained. The percentage of spurious
matches is estimated to be about 1\% (BSC) and about 5\% (FSC)
(i.e. about 3 - 4 sources in total). 

In total we have obtained 80 matches with the BSC plus 48 matches with
the FSC, for a total of 128 matches. Considering only the blazars
candidates (as defined in Section~4), 119 have been detected in the
RASS catalogue. The percentage of X-ray detections is thus slightly
higher (45\%) among the blazars candidates (i.e. core-dominated
sources) than among the total sample (42\%). Only about 30\% of the
low-power blazar candidates ($P_{5} \leq$10$^{25}$ W Hz$^{-1}$) have been
detected in the RASS.

The X-ray fluxes have been computed from the published count-rates by
assuming a photon index of 2 ($\alpha_X$=1) and de-absorbed for the
galactic absorption by using the galactic column density presented in
Stark et al. (1992). For the sources not detected in the RASS
catalogues we need an estimate of the upper limit on the X-ray
flux. This is not a simple task since the RASS does not have a
``flat'' sky coverage and the actual count-rate limit depends on the
sky position.  
We have estimated a ``statistical''
upper limit in the following way.  We have performed a positional
cross-correlation of the RASS catalogues with a deeper X-ray
catalogue, based on the pointed ROSAT PSPC images, used to define the
REX survey (see Caccianiga et al. 1999 for the details). This
catalogue is based on the analysis of 1202 pointed ROSAT PSPC images
and reaches flux limits of $\sim$3-5$\times$10$^{-14}$
erg\,s$^{-1}$\,cm$^{-2}$ in the 0.5-2.0 keV energy range on a sky
region of 2183 deg$^{2}$.  We have only considered the sources in the
area covered by the CBS sample to check the sensitivity of the RASS in
the right sky position, and those detected with a high significance
($\sigma>$10, see Caccianiga et al. 1999).  We have then evaluated the
count-rate above which more than 90\% of the sources included in the
deep X-ray catalogue have a counterpart in the RASS catalogue. We have
used this count-rate as an upper limit for the non-detection.  We note
that this value (CR=0.05 counts s$^{-1}$) is higher than the faintest
count-rates found in the RASS FSC, which reach values as faint as 0.01
counts s$^{-1}$.

The upper-limits of the X-ray fluxes were then computed by assuming this
conservative limiting count-rate of 0.05 counts s$^{-1}$, and
converted into a de-absorbed X-ray flux on the basis of the galactic
column density. 

\begin{figure} 
\centerline{ 
\psfig{file=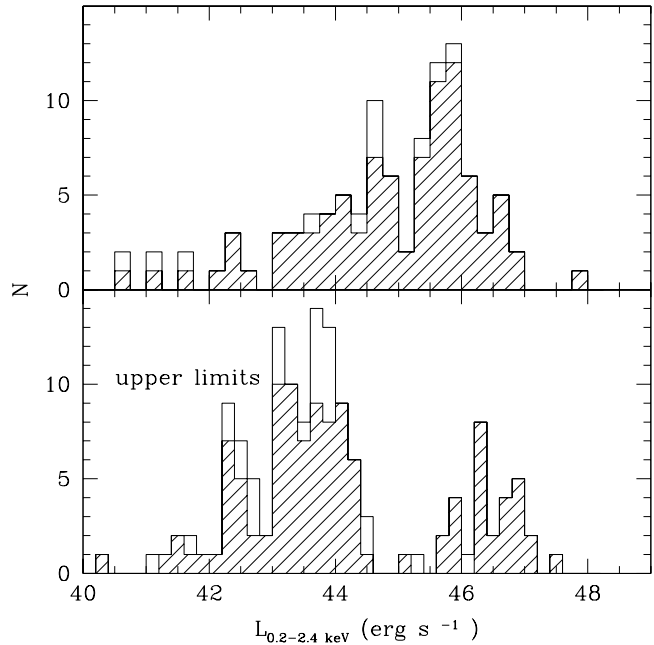,height=9cm,width=9cm}
} 
\caption{{\it upper panel}: The distribution of the X-ray luminosities of the CBS
sources detected in the RASS survey. The shaded area indicates the
core-dominated objects  as defined in the text. {\it Lower panel:} The 
distribution of the upper limits on the X-ray luminosities for the objects
not detected in the RASS survey. The shaded area indicates the
core-dominated objects}
\label{lx} 
\end{figure}

Figure \ref{lx} shows the X-ray luminosity distribution for the
CBS sources.
The X-ray fluxes are used in the next section to estimate the
radio-to-X-ray luminosity ratio of the sources in the CBS sample 
detected in the RASS catalogues.

\section{The $\alpha_{\rm{rx}}$ distribution and the blazar sequence}

According to the blazar sequence, the low-power regime should be
dominated by high-energy peaked sources, characterized by small (flat)
values of the radio-to-X-ray spectral index
($\alpha_{RX}\leq$0.75). 

\begin{figure} 
\centerline{ 
\psfig{file=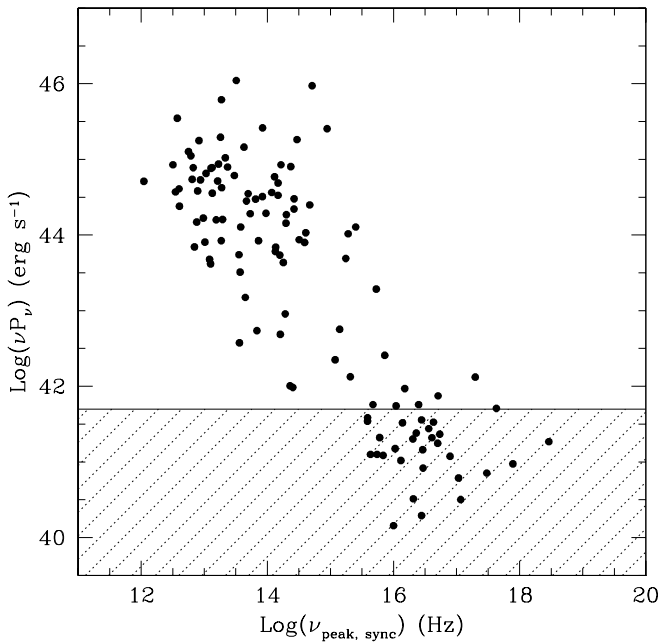,height=9cm,width=9cm}
} 
\caption{The radio power (multiplied by the frequency) in erg s$^{-1}$
versus the synchrotron peak frequency (in Hz)  of the blazars 
considered by Fossati et al. (1998) to derive the ``blazar sequence''.
According to this sequence, all the blazars with a radio power
below 10$^{25}$ W Hz$^{-1}$ should have a synchrotron peak frequency larger
than $\sim$10$^{15.5}$ Hz (shaded area). Adapted from Fossati et al. (1998). }
\label{foss_seq} 
\end{figure}

\begin{figure} 
\centerline{ 
\psfig{file=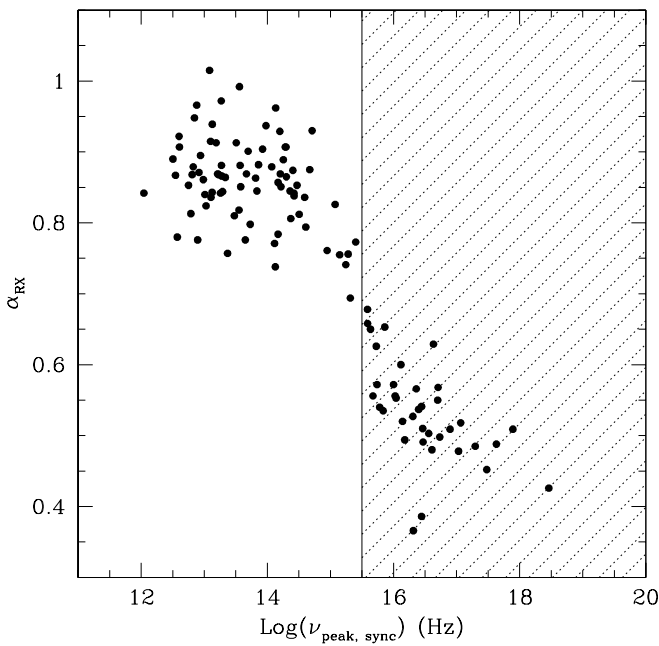,height=9cm,width=9cm}
} 
\caption{The $\alpha_{RX}$ values versus the
synchrotron peak frequency for the blazars studied by Fossati et al. (1998). 
The expected values of $\alpha_{RX}$ for blazars with a radio
power below 10$^{25}$ W Hz$^{-1}$ (shaded area, see previous Figure) 
are in the range between 0.35 and 0.7. 
Adapted from Fossati et al. (1998)
}
\label{arx_foss} 
\end{figure}

To better quantify this statement, we have reproduced here two figures
(Figure~\ref{foss_seq} and Figure~\ref{arx_foss}) taken from Fossati
et al. (1998).  The blazar sequence as defined by Fossati et
al. (1998) is reported in Figure~\ref{foss_seq}. According to this
sequence, all the blazars with a radio power below 10$^{25}$ W
Hz$^{-1}$ (corresponding to $\nu$L$_{\nu}$=10$^{41.7}$ erg s$^{-1}$)
are expected to have a synchrotron peak frequency larger than
$\sim$10$^{15.5}$ Hz. If we now look at Figure~\ref{arx_foss} we see
that this high peak frequency translates into a very flat
$\alpha_{RX}$ value. Although the correlation between $\alpha_{RX}$
and the synchrotron peak frequency has  quite a large scatter, it is
true that all the sources considered by Fossati et al. (1998)
with a synchrotron peak frequency larger
than $\sim$10$^{15.5}$ Hz have a $\alpha_{RX}$ in the range 0.4-0.75.  
This statement holds also using the
blazars in the DXRBS sample (see Figure~11 from Padovani et al. 2003),
although the number of objects in this sample with a synchrotron peak
frequency larger than $\sim$10$^{15.5}$ Hz is very small (3 sources in
total)\footnote{The lack of objects in this region of the parameter
space is an obvious consequence to the fact that the DXRBS is not very
sensitive to the extreme HBLs so the high peak frequency range
(between 10$^{16}$ and 10$^{18}$ Hz) is poorly sampled.}

Thus, we expect that all the blazar candidates in the
10$^{23}$-10$^{25}$ W Hz$^{-1}$ power range have an $\alpha_{RX}$
flatter than 0.7-0.75, if they fit the blazar sequence.

We note that such a flat $\alpha_{RX}$ value would correspond to X-ray fluxes
larger than 1.5$\times$10$^{-13}$ erg s$^{-1}$ cm$^{-2}$ (0.5-2.0 keV) 
given the radio fluxes of the CBS sources (S$_{5 GHz}>$30 mJy). 
Thus, many of the low-power blazar candidates should have been detected in 
the RASS survey. Instead, as described in the previous section,
only 30\% of the low power objects have been detected in the RASS
survey.

%In Figure~\ref{samples} the values of $\alpha_{RX}$ computed for the
%blazars in different samples taken from the literature, are plotted
%against the radio power. In particular, data from the 2 ``classical''
%BL Lac samples, i.e. the EMSS (Morris et al. 1991) and 1~Jy sample
%(Stickel et al. 1991), are used together with the more recent 
%samples, i.e  the RGB (Laurent-Muehleisen et al. 1998), 
%the DXRBS (Perlman et al. 1998) and the XB-REX 
%(Caccianiga et al. 2002). We note that the DXRBS 
%includes both BL Lacs and FSRQ.  The presence of a correlation
%between the two quantities is evident. Specifically, the two lines
%plotted enclose more than 95\% of the blazars in those samples.

\begin{figure}
\centerline{
\psfig{file=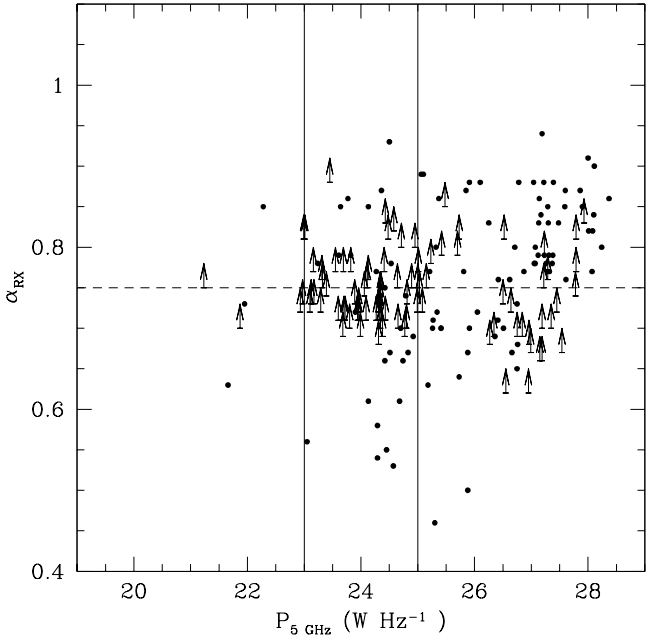,height=9cm,width=9cm} 
}
\caption{Two-point spectral index ($\alpha_{RX}$) versus 
the radio power at 5 GHz for the CBS blazar candidates. 
For the objects not 
detected in the RASS survey the lower limits on $\alpha_{RX}$ are indicated
(see text for details). The two continuous lines define the range of radio 
power 
where, according to the blazar sequence, we should have found only blazars
with ``flat'' ($<$0.75) $\alpha_{RX}$ values. The dashed line indicates 
$\alpha_{RX}$=0.75} 
\label{power_arx} 
\end{figure}

In Figure~\ref{power_arx} we have plotted the values of $\alpha_{RX}$
for the CBS blazar candidates including the lower-limits on
$\alpha_{RX}$ for the sources not detected in the RASS.  The objects
without spectroscopic classification, and the BL Lacs without redshift
have been excluded from the plots.

The important conclusion which can be drawn from the analysis of
 Figure~\ref{power_arx} is that a large fraction of the objects
in the low power regime show $\alpha_{RX}$ values (or lower limits) 
much steeper than 0.75. 
 
%many sources in the top left part of the plot which 
%have no counterpart in Figure~\ref{samples}.

\begin{figure} 
\centerline{
\psfig{file=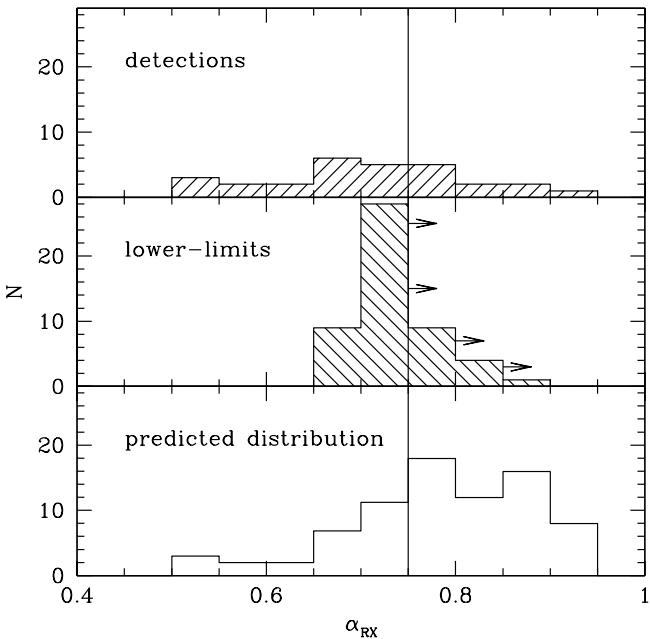,height=9cm,width=9cm} 
}
\caption{The two-point spectral index ($\alpha_{RX}$) distribution 
for the sources in the CBS blazar candidates  and with a 
radio power between 10$^{23}$ and 10$^{25}$ W Hz$^{-1}$. {\it top panel}: 
sources detected in the RASS; {\it middle panel}: sources not
detected in the RASS; {\it lower panel}: the expected ``real'' 
distribution computed with the method described in Avni et al. (1980)
(see text for details)} 
\label{arx} 
\end{figure}

More quantitatively, we have considered only the sources 
with a radio power between 10$^{23}$ and 10$^{25}$ W Hz$^{-1}$
and we have plotted the distribution of the corresponding $\alpha_{RX}$ values 
(Figure~\ref{arx}). If we consider the lower limits 
as actual detection, the fraction of sources with $\alpha_{RX}$
steeper than 0.75 is 30\%. This should be considered as a lower
limit since the non detections are expected to be distributed
at larger values of $\alpha_{RX}$. By using the non-parametric
method described in Avni et al. (1980), which 
provides an analytic solution for the best estimate 
of a distribution function of one (binned) independent variable
(the $\alpha_{RX}$) taking into account
the lower limits, we have derived the
expected real $\alpha_{RX}$ distribution  (lower panel in Figure~\ref{arx}).
Based on this distribution, the percentage of
sources in the low radio power range with $\alpha_{RX}>$0.75 
is 68\%. 

It is worth noting that for a number of objects detected in
the RASS, the X-ray emission could be partly due to the
presence of hot gas associated to the host galaxy. Since the thermal
X-ray luminosity of elliptical galaxies can reach values up
to 10$^{42}$-10$^{43}$ erg s$^{-1}$ (Forman et al. 1994),
only the few sources in the low luminosity tail of the distribution
presented in Figure~\ref{lx} can be affected by this 
problem. In any case, after having taken into account the
thermal contribution from the host galaxy, the corrected 
values of $\alpha_{RX}$ are expected to be steeper than 
uncorrected ones, thus populating even more the steep $\alpha_{RX}$ 
region of Figure~\ref{power_arx}.

We conclude that a large fraction (between 30\% and 68\%) of the 
core-dominated sources selected in the CBS with a radio power between 
10$^{23}$ and 10$^{25}$ W Hz$^{-1}$ have $\alpha_{RX}$ values
steeper than 0.75, thus they lie outside the blazar sequence.

In the next sections we investigate 3 possible explanations
for the large spread of the $\alpha_{RX}$ values observed at low-powers:

\begin{itemize}
\item The sources found outside the predicted sequence are
spurious optical identifications (Section 8);

\item At low-powers the orientation effects become important and some weakly
beamed but intrinsically high-power objects are selected (Section 9).

\item For powers below 10$^{25}$ W Hz$^{-1}$ a population
of sources different from blazars and characterized by a 
weak (or even absent) X-ray emission, is getting into the sample 
thus spreading the blazar sequence (Section 10);

\end{itemize}

%In particular, we study any possible contamination (spurious radio/optical
%matches, non-blazar sources) which can artificially spread the
%blazar sequence. 
%We will show that a number of non-blazars (CSO) are actually
%expected in the CBS and that only with high resolution data (VLBI) it
%will be possible to separate these objects from the real blazars.
%However, with the data available so far we have extracted at least
%2 sources for which the  blazar nature has been convincingly assesed
%and for which the low-power and the steep $\alpha_{RX}$ values
%do not fit the blazar sequence.
%Finally we discuss the possibility that the blazar sequence is
%a result of a biased sampling of the '$\alpha_{RX} - P_{5}$' plane. 

\section{Spurious identifications?}

Among the 24 low-power sources (P$_{5 GHz}$=10$^{23}$-10$^{25}$ W
Hz$^{-1}$) with $\alpha_{RX}>$0.75 (or the lower limit on the
$\alpha_{RX}>$ larger than this value) there are 10 galaxies with no
obvious signs of activity in their optical spectra (6 have the typical
spectrum of an elliptical galaxy and 4 have narrow emission lines
probably associated with the host galaxy), and another 10 objects
showing a broad H$\alpha$ emission line.  The remaining 4 sources are
classified in the literature as Sy2/LINERS due to the presence of
narrow emission lines in the optical spectrum.  In principle, some of
the ``normal'' galaxies could be objects included by mistake in the
cross-correlation with the optical catalogue (APM\footnote{\it
www.ast.cam.ac.uk/$\sim$apmcat/}), while the real counterpart of the
radio source could be a fainter (more distant) blazar. In this case,
the redshift (and thus the radio power) would be underestimated while
the $\alpha_{RX}$ value would be the one relative to the powerful
blazar (i.e. a steep $\alpha_{RX}$, according to the blazar
sequence). This could explain, in principle, the peculiar position of
these sources in the $\alpha_{RX}$/P$_{5_{\rm GHz}}$ plane.

In Paper~I we have described in detail the technique used to find the
optical counterparts of the radio sources in the CLASS sample. In
order to assess the reliability of that identification procedure we
have run ten times the positional cross-correlation between the CLASS
and the APM catalogues, after having applied a positional offset to
the CLASS sources from $\Delta\delta$=+1$\arcmin$ to +10$\arcmin$,
plus an additional cross-correlation with $\Delta\delta$=+1 degree. In
this way, all the matches are expected to be of spurious origin. After
having applied the same selection criteria used to select the CBS
sample, we have found 12 spurious point-like sources (APM
classification equal to ``-1''), plus 3 galaxies (APM classification
= ``1''), and 4 blends (APM classification = ``2'').  So far, 5
objects have been spectroscopically identified as stars and they
probably make up the point-like spurious matches.  The case of
spurious galaxies, however, is clearly a less simple one, as these
represent credible counterparts of the radio source. Nevertheless, the
fact that we only expect $\sim$3-7 (including the blends) 
spurious matches means that this would
be insufficient to explain the deviation from the 'blazar
sequence'. In any case, we performed another check by plotting the
very accurate ($\sim$arcsec) 8.4 GHz positions taken with VLA A-array
on the optical finding chart of some of the 24 sources with a steep
$\alpha_{RX}$ in the low power regime. We found  that in all cases the
radio position is consistent with the optical nucleus thus confirming that
the radio and the optical emission come from the same source.

Hence, the spurious radio/optical matches do not explain 
the discrepancy observed in the $\alpha_{RX}$/P$_{5 GHz}$ plane.

\section{Orientation effects?}

\begin{figure} 
\centerline{
\psfig{file=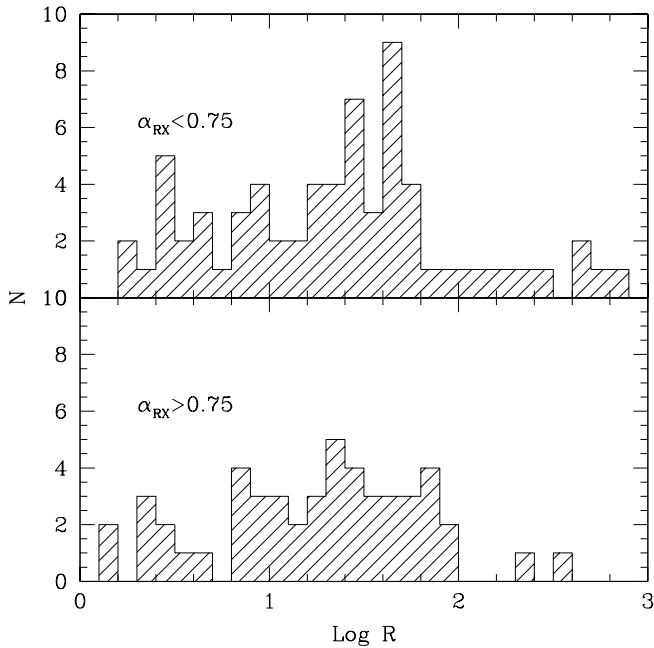,height=9cm,width=9cm} 
} 
\caption{The core-dominance parameter distribution for the 
blazars with P$_{5GHz}<$10$^{25}$ WHz$^{-1}$: with  
$\alpha_{RX}<$0.75 (upper panel), and with 
$\alpha_{RX}>$0.75 (bottom panel). } 
\label{r_arx} 
\end{figure}

In this Section we investigate the possibility that orientation
effects are responsible for the deviation from the blazar sequence
found in the CBS sample.

If an intrinsically high-power blazar, with a steep $\alpha_{\rm{rx}}$,
like those observed in the 1~Jy sample, is observed at larger angles, 
the Doppler factor ($\delta$) 
decreases  with the observing angle $\theta$ and the observed radio power 
is expected to decrease rapidly, i.e. proportionally 
to $\delta^p$, with p=2+$\alpha\sim$2.4 (see Urry \& Padovani 1995). 
The observed $\alpha_{RX}$, instead, 
changes only if the radio and the X-ray spectral indices are
different (see, for instance, Chiaberge et al. 2000). 
Therefore, 
a mis-alignment brings the object toward the low-power section of 
Figure~\ref{r_arx} and toward higher or lower values of $\alpha_{RX}$ 
depending on the actual radio and X-ray indices of the source. 
As a consequence, a mis-alignment jet/observer can spread the blazar sequence 
and, in particular, it may populate the low-power/steep $\alpha_{RX}$ region.

The possibility that the spread in $\alpha_{\rm{rx}}$ observed at low radio
powers is due to some orientation bias can be tested by analysing the
radio properties of the CBS sources. In particular, if this hypothesis
is correct, we would expect low-power/high-$\alpha_{RX}$ sources to be
less core-dominated than the others.

More quantitatively, the distribution of the core-dominance parameter
for the sources with radio power below 10$^{25}$ W Hz$^{-1}$ is shown
in Figure~\ref{r_arx}. The top panel shows the distribution of the
sources with $\alpha_{RX}<$0.75, and the bottom panel shows the
distribution of those sources with $\alpha_{RX}>$0.75.  The two
distributions are not significantly different (Kolmogorov-Smirnov
probability = 75\% for the null hypothesis).

\begin{figure} 
\centerline{
\psfig{file=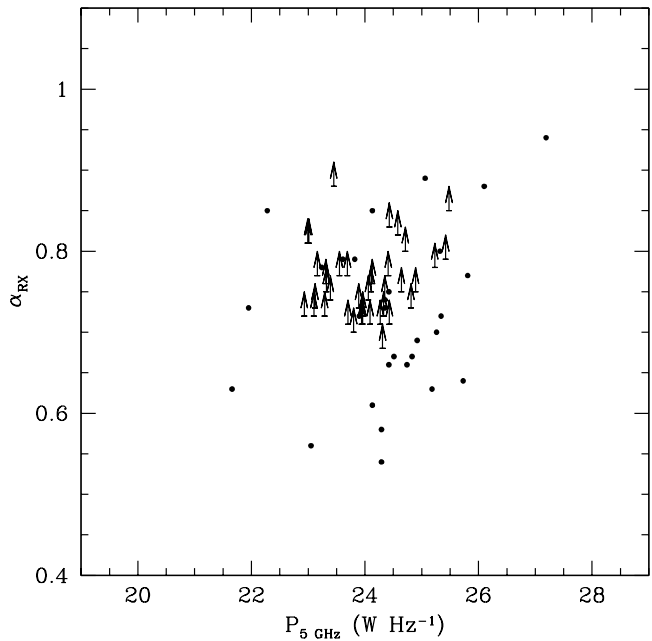,height=9cm,width=9cm} 
} 
\caption{Two-point spectral index ($\alpha_{RX}$) versus 
the radio power at 5 GHz for the sources in the CLASS Blazar
Survey for the sources with z$<$0.3 and with 
R$>$10.} 
\label{power_arx_coredom} 
\end{figure}

In Figure~\ref{power_arx_coredom} we have re-produced
Figure~\ref{power_arx} using only the blazar candidates with z$<0.3$ and
R$>$10. If the orientation was really the cause for the observed
spread at low-power of Figure~\ref{power_arx}, we would expect that by
using only the highly core-dominated sources, the resulting plot would
show a better power/$\alpha_{RX}$ correlation. This, however, is not
the case, and we find that the sources with low-power and
high/$\alpha_{RX}$ are still present. This result suggests that
orientation is not the main reason for the big spread of $\alpha_{RX}$
values observed at low radio powers.

In Figure~\ref{map} we show the radio maps (VLA B-array at 1.4~GHz)
for the 24 low-power blazar candidates that do not fit the
blazar sequence, i.e.  with $\alpha_{RX}>$0.75.
Given the redshift range covered by these 24 sources (0.01
-- 0.1) the resolution of the maps corresponds to linear scales
between 1.4 kpc and 14 kpc.

\section{Contamination by non-blazars}

In this Section we investigate the possibility that the spread in the
'$\alpha_{RX}$/P$_{5 GHz}$' plane found in the CBS is the result of
contamination by sources that are not true blazars but which slipped
in the sample because of the selection criteria. These non-blazars
must  have the following properties:

%The radio compactness (at arcsecond scale) used to 
%classify a source as blazar candidate is not  
%100\% reliable to select true blazar and a 
%number of non-blazar sources are expected in the CBS sample. 
%In particular, in the low-power regime different
%types of sources can be selected and contribute to the
%spread observed in the $\alpha_{RX}$/P$_{5 GHz}$ plot.
%The main characteristics of this population would be:

1) weak X-ray emission when compared to the radio emission;

2) radio power at 5~GHz between 10$^{23}$ and 10$^{25}$ WHz$^{-1}$ 
with only few cases of sources with a radio power below 
10$^{23}$ WHz$^{-1}$;

3) flat-spectrum and core-dominated (at arcsecond resolutions).
 
Overall this population must have 5~GHz radio powers consistent with
those of the HBLs typically selected in X-ray survey (Slew Survey,
EMSS) but with X-ray luminosities more than 10 times lower. 
%Furthermore, the fact
%that show flat radio spectrum and core dominated make 
%the classification of these objects as ``normal'' (i.e. unbeamed) 
%radio galaxies quite improbable (see Section~4).

%The mean radio power of these sources ($\sim$10$^{24}$ WHz$^{-1}$) is
%consistent with the 5~GHz low-power range of FR~I radio galaxies 
%(see for instance Urry \& Padovani 1995). 
%However, the observed radio compactness and the flat spectra make 
%the classification of these objects as ``normal'' (i.e. unbeamed) 
%radio galaxies quite improbable (see Section~4).
%The range of radio power is the same observed in the typical HBLs
%selected in the EMSS or in the Einstein Slew Survey, but the 
%X-ray luminosity is more than a factor 10 lower than that observed in these
%HBLs.

Different possibilities are analysed below.

\subsection{GPS/CSO objects}
A viable possibility is that these sources are Gigahertz Peaked
Sources (GPS) or Compact Symmetric Objects (CSO). These sources are
compact at these resolutions and, in many cases, they appear as flat
spectrum objects between 1.4 and 5 GHz. A number of CSO/GPS are
actually expected in the CLASS survey. Current data suggests that the
percentage of CSO/GPS sources in radio selected samples varies from 2
to 11\% (Peck \& Taylor 2000).  The problem, however, resides in
estimating the percentage of this type of sources at the powers 
sampled by the CBS. In fact, the
Luminosity Function (LF) of CSO/GPS sources is very badly determined,
particularly at radio luminosities below 10$^{25}$ WHz$^{-1}$ (see for
instance, Snellen et al. 2000). 

In order to address this issue we make use of current investigations
concerning the 200~mJy sample (March\~a et al. 1996), which has
similar selection criteria to the CBS and for which VLBI data exist
for the majority of the sources. The 200~mJy sample has $\sim$60 flat
radio spectrum sources with flux density limit of 200~mJy, and optical
selection similar to that of the CBS. VLBI observations of the 200~mJy
sample do indeed find a number of CSOs sources (Polatidis et al. in
prep). An '$\alpha_{RX}$-P$_{5 GHz}$' plot for the 200~mJy sample
yields a similar excess of sources to the one found in the CBS ($\sim
30$\%) above the blazar sequence for radio powers $<
10^{25}$WHz$^{-1}$. Making use of the VLBI data currently available we
find that among the eleven sources outside the sequence, three 
have a CSO morphology, while the remaining show a core-jet morphology,
as expected in the case of blazars.  At present we do not have similar
results for the CBS, but radio observations at higher (VLBA)
resolutions of some of these sources are underway.  However, if the
percentage of the CSOs is the same as in the 200~mJy sample, these
sources cannot represent a completely satisfactory explanation for the
observed excess of steep $\alpha_{RX}$ objects in the CBS.

\subsection{The high luminosity end of radio-quiet sources}

Another possible population which can be present in the CBS 
sample is that of the radio-quiet  AGNs (i.e. Seyfert galaxies) whose 
radio powers range from 10$^{18}$ to 10$^{24}$ W Hz$^{-1}$ 
(e.g. Ulvestad \& Wilson 1989; Ulvestad \& Ho 2001a). It must be
considered, however, that the majority ($\sim$90\%, Ulvestad \& Wilson 1989) 
of powerful Seyferts have steep radio spectra and they should not be 
efficiently selected in the CBS sample. An important exception is that 
of Low-luminosity Seyeferts:  the analysis of the radio properties of 
45 low-luminosity Seyfert galaxies (type 1 and type 2) taken from the 
Palomar spectroscopic survey of
nearby galaxies undertaken by Ulvestad \& Ho (2001a) has shown that a
significant fraction ($\sim$50\%) of them have flat or inverted radio
spectra. The origin of the flat radio radio emission in these sources could 
be related to a jet-like (but not necessarily beamed) structure, to 
an ADAF activity or to a combination of the two mechanisms (see for instance 
Ulvestad \& Ho 2001b). 

The radio powers observed in these AGNs are typically below 10$^{23}$ W
Hz$^{-1}$ and, even if the 8 objects in Figure~\ref{power_arx} with a 
radio power below 10$^{23}$ W Hz$^{-1}$ were potentially 
similar to the Seyfert galaxies
studied by Ulvestad \& Ho\footnote{indeed, one object is in common, namely
the source GB6J065010+605001 (NGC~2273)}, the mean radio power 
of the 24 sources under study here (10$^{24}$ W Hz$^{-1}$) is 4 orders of
magnitude larger than the mean radio power of the Seyfert galaxies
from the Palomar spectroscopic survey (10$^{20}$ W Hz$^{-1}$).
Therefore, the contamination by this kind of objects in the CBS sample
must be very low and should affect only the very low-power range 
($<$10$^{23}$ W Hz$^{-1}$).

\subsection{ULIRG}

Among the 24 objects that are 'out of the sequence', there are two
well known Ultra Luminous Infrared Galaxies (ULIRG), namely Mrk 231
and Mrk 273, in which the extremely intense starburst activity
contributes to a significant fraction of the radio power
(e.g. Ulvestad, Wrobel \& Carilli 1999; Carilli \& Taylor 2000). In
the case of Mrk 231 the nuclear radio component not related to the
starburst activity (about 100~mJy) could be associated to a CSO
(Lonsdale et al. 2003).  The SEDs of these 2 sources are dominated by
the far-IR emission (detected with IRAS).  Among the 24 sources there
are 3 more objects which are detected in IRAS but with a FIR
luminosity below the 10$^{12}$ L$\odot$ and are not classified as
ULIRG.

Except for the two ULIRG Mrk 231 and Mrk 273, in the other 
objects the starburst emission (if present)  is not 
expected to be the dominant component.
In any case, accurate VLBI measurements
are needed in order to separate the emission due to the starburst 
phenomenon from the genuine AGN emission. A similar analysis must be
done also in the  X-ray band. 

%The SEDs of these 2 sources are clearly different from the
%SEDs of the remaining 22 objects as they are dominated by the far-IR
%emission (detected with IRAS).
%These are the only 2 ULIRG present in the sample
%and they should be flagged as ``non-blazar". 
%In any case, the SEDs of these 2 sources are clearly different from the
%SEDs of the remaining 22 objects as they are dominated by the far-IR
%emission (detected with IRAS).
% and it is unlikely that similar sources
%are present in the CBS sample.

\subsection{Conclusion about the contaminating sources}

From the discussion above we can conclude that some contamination from
non-blazars is expected in the CBS, something that indicates that also
a selection criterion based solely on the radio compactness (at the VLA
resolution) may not be 100\% reliable in selecting blazar, in
particular in the low-power regime. Clearly, higher resolution data
(VLBI) will be needed to disentangle the non-blazars from the blazar
population. For this reason a systematic observation of the CBS
sources with VLBI is in progress.  At present, however, we already
have 2 good confirmations of low power blazars which defy the proposed
'blazar sequence':

{\it GB6J022526+371029 (CGCG523-037)}. This object (z=0.0334) shows an
optical spectrum dominated by the host galaxy with a few emission
lines (H$\beta$, [OIII]$\lambda$5007\AA, [OI], H$\alpha$, [NII],
[SII]) (Caccianiga et al. 2002). It belongs also to the B2 catalogue
and it has been classified as Low-power compact sources (LPC,
Giovannini et al. 2001) due to its low-power and high compactness. The
object has been observed with the European VLBI Network (EVN) and with
the VLBA by Giovannini et al. (2001). In both observations an
unresolved source has been detected with a total flux density
comparable to the arcsecond core flux density. Based on these
observations, Giovannini et al. (2001) suggested for this object a
classification as low-power (P$_{5 GHz}$=7$\times$10$^{23}$ W Hz$^{-1}$) 
BL Lac whose observed core power is too
low to dominate the optical emission. The non-detection in the RASS
implies a $\alpha_{RX}>$0.77.
%Based on ROSAT HRI observations, Canosa et al. (2003)
%has derived a more stringent upper limit on the X-ray core flux which
%implies $\alpha_{RX}>$0.9. CHECK!!

{\it GB6J164734+494954 (JVAS J1647+499)}. The blazar nature of
this source (z=0.0475) has been clearly confirmed by the detection 
of a high and variable optical polarization (Jackson \& March\~a 1999). 
The VLBA map at 4.9~GHz (Bondi et al. 2001)  shows
a bright core plus a collimated one-sided jet. In the X-ray, 
the source has been detected in the RASS with a count rate of 0.097 counts/s
corresponding to an unabsorbed X-ray flux in the 0.1-2.4 band of
1.4$\times$10$^{-12}$ erg s$^{-1}$ cm$^{-2}$ (assuming a power law
spectrum of $\Gamma$=2). The corresponding $\alpha_{RX}$ is 0.75-0.77,
using the core flux at 4.9 GHz published by Bondi et al. (2001) or
the GB6 flux respectively. The radio power at 5~GHz is 1.9$\times$10$^{24}$ 
W Hz$^{-1}$.

The existence of these 2 low-power blazars with a steep $\alpha_{RX}$ 
value (thus suggesting a synchrotron peak frequency below 10$^{15.5}$ Hz) 
constitutes confirmation that there is indeed a deviation from the proposed blazar
sequence. What remains to be established is the quantification of this
deviation, something that will require the collection of more data.

\section{Summary and conclusions}

In this paper we have analysed the radio and X-ray data of the CLASS Blazar
Survey sample which is, until now, the best radio selected
sample to study blazars in the low radio power regime. 
In order to avoid any bias introduced by the optical classification
in this luminosity regime, we have used the radio data to 
classify an object as blazar candidate. Specifically we
have used VLA data at 1.4~GHz and B-array (FWHM$\sim$5$\arcsec$), and
at 8.4~GHz and A-array (FWHM$\sim$0.24$\arcsec$) to constrain the
actual core dominance of the sources included in the CBS. Based on
these data, we have defined as blazar candidates those sources
which are core-dominated (R$>$1). We have then 
used the X-ray information derived from the ROSAT All Sky Survey to
compute the radio-to-X-ray flux ratios ($\alpha_{RX}$) of the 
blazar candidates in the sample. The main goal
of this study was to test the so called ``blazar sequence'' 
proposed to unify the blazar class (Fossati et al. 1998; Ghisellini
et al. 1998). According to this model, and in the range of radio
powers under study, we should find only High-energy peaked BL Lac
(HBL), i.e. blazars whose synchrotron emission extends up to the
X-rays, producing a high X-ray-to-radio flux ratio ($\alpha_{RX}<0.75$). 
This is not what we find. The main results
of the study can be summarised as follows:

\begin{itemize}

\item In the 10$^{23}$-10$^{25}$ WHz$^{-1}$ radio power regime, a
large fraction (between 30\% and 68\%) of the sources have a steeper (larger)
$\alpha_{RX}$ than what is expected from the blazar sequence 
traced by previous BL Lac/blazars samples;

\item The deviation from the blazar sequence found in the sources of
the CBS is unlikely to be due to orientation effects. This is
supported by the fact that the deviation from the sequence is still
observed when we consider only those sources which are strongly radio
core-dominated (R$>$10) at a linear scale of $\leq$30 kpc;

\item The possibility that contaminating sources could be at the
origin of the observed deviation from the 'blazar sequence' was also
investigated. The most likely candidates for this contamination are
the Compact Symmetric Sources (CSOs) that also appear as compact
radio sources at these spatial resolutions. However, we conclude that,
if the percentage of CSOs in the CBS is similar to that found in other
samples, then it is insufficient to justify the observed deviation. 
VLBA observations are in progress to distinguish core-jet from 
double morphology. 
%Other 2 peculiar sources, namely the ULIRG Mrk231 and Mrk273, are also
%present among the sources that do not fit the blazar sequence.

\item Based on the data available so far we have extracted 2 sources
out of the 24 objects that do not fit the blazar sequence, with
a core-jet or compact radio morphology at the m.a.s. resolution,
thus ruling out the CSO nature.

\end{itemize}

One of the prime motivations for the selection of the CBS was widening
the range of blazar parameter space sampled. Indeed, due to its low
radio flux density limit and no X-ray selection, the CBS has made
it possible to find sources at low radio powers and with 
steep $\alpha_{RX}$, i.e. outside the expected sequence across that plane. 
Even if the contamination by sources other than blazars can be important
at these luminosities we have found at least two good cases where the
blazar nature of the source is suggested from high resolution (VLBI) maps.
We stress here that this kind of source is heavily undersampled in current
surveys sensitive to low-power blazars because of the presence, in these
surveys, of relatively bright X-ray flux limits.

This result suggests that the 'blazar sequence' could be, at least
partly, due to selection effects resulting from poor sampling of the
relevant parameter space, thus confirming recent results from the DXRBS 
survey (Padovani et al. 2003).

According to the models which unify the blazar class,
the observed peak of the emission is determined by the cooling processes 
at work (e.g. Ghisellini, Celotti \& Costamante 2002). 
While the existence of an intrinsic physical limit
to the synchrotron peak frequency for the most powerful 
blazar seems to be supported by recent work 
(Padovani et al. 2003), the large spread in the 
$\alpha_{RX}$ values observed 
in the CBS sample may suggest that in the low-power 
regime different levels of cooling, and thus
of peak frequency, may be present. 
The physical reason of this spread must be 
investigated.
It has been noted by Ghisellini et al. (1998) 
that the increasing importance of an external radiation field due 
for instance to the presence of the broad emission lines could
increase the cooling level dictating in this way the peak
energy of the emitting particles. 
Therefore, it will be interesting to test whether the observed 
$\alpha_{RX}$ values are correlated with the optical
properties of the sources and, in particular, with the presence and
intensity of the emission lines. As already discussed in Section 3, a careful
analysis of the nuclear optical spectra at these power regimes 
can be done only by 
properly taking into account the presence of the host galaxy. 
This will be the subject of a forthcoming paper.

%Overall, the result of the current investigation puts in evidence the
%importance of sampling the blazar phenomenon in the low power regime,
%where the unifying model represented by the 'blazar sequence' may no
%longer be a faithful description.

\section*{Acknowledgments} 
%This research was supported by the European
%Commission Training and Mobility of Researchers program, research
%network contract ERBFMRX-CT96-0034 ``CERES''.  
We would like to thank I.W.A. Browne and an anonymous referee 
for useful comments and discussions and G. Fossati, for providing us 
with the data on the blazar sequence. 
This research has received partial financial 
support from the Portuguese Funda\c{c}\~ao para a 
Ci\^encia e Tecnologia (FCT, PRO 15132/1999 and SFRH/BPD/3610/2000).

\begin{onecolumn}

\begin{figure}
\begin{tabular}{c c}
 GB6J022526+371029 & GB6J061641+663024 \\
\psfig{figure=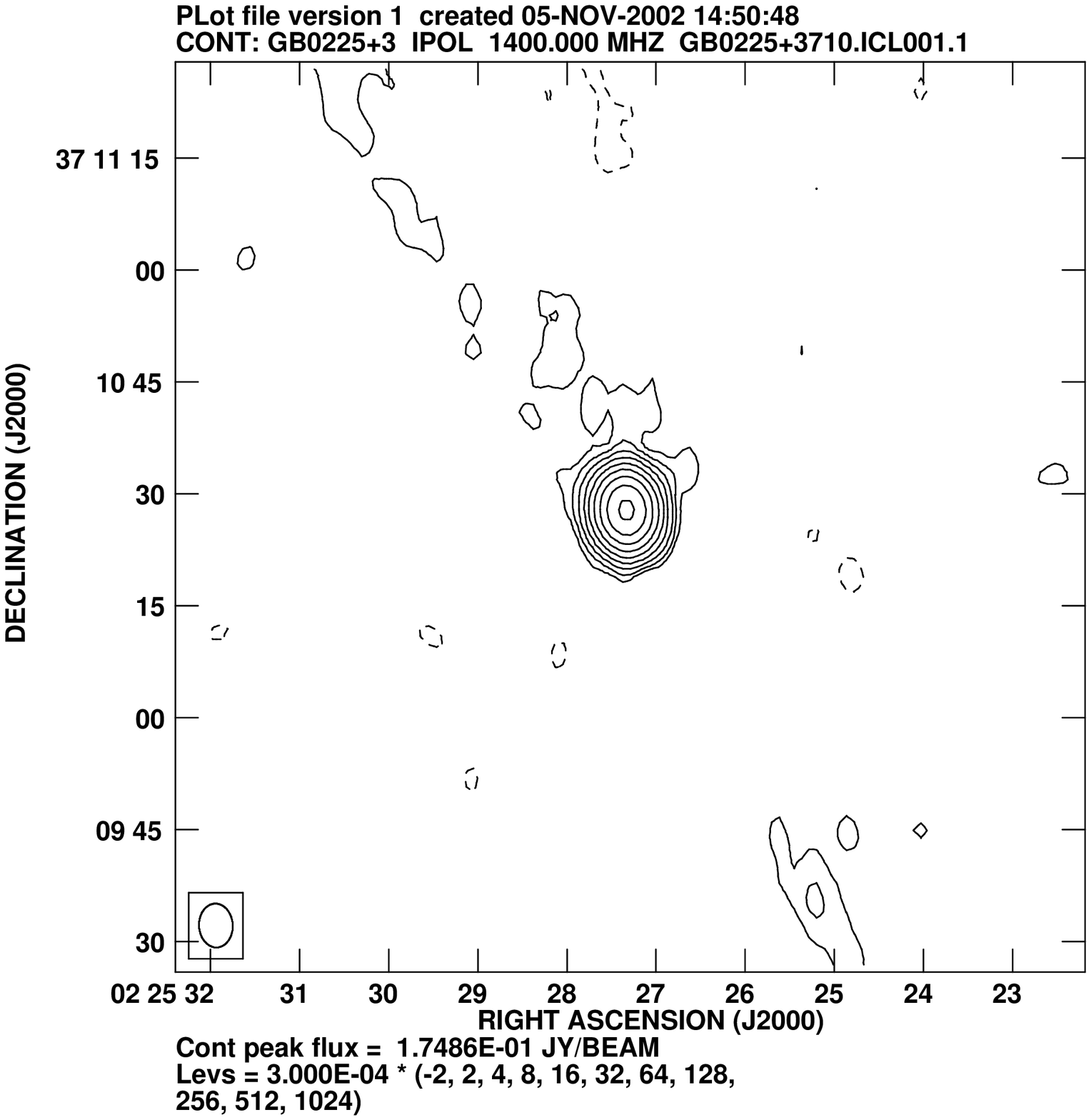,height=7cm,width=7cm} &
\psfig{figure=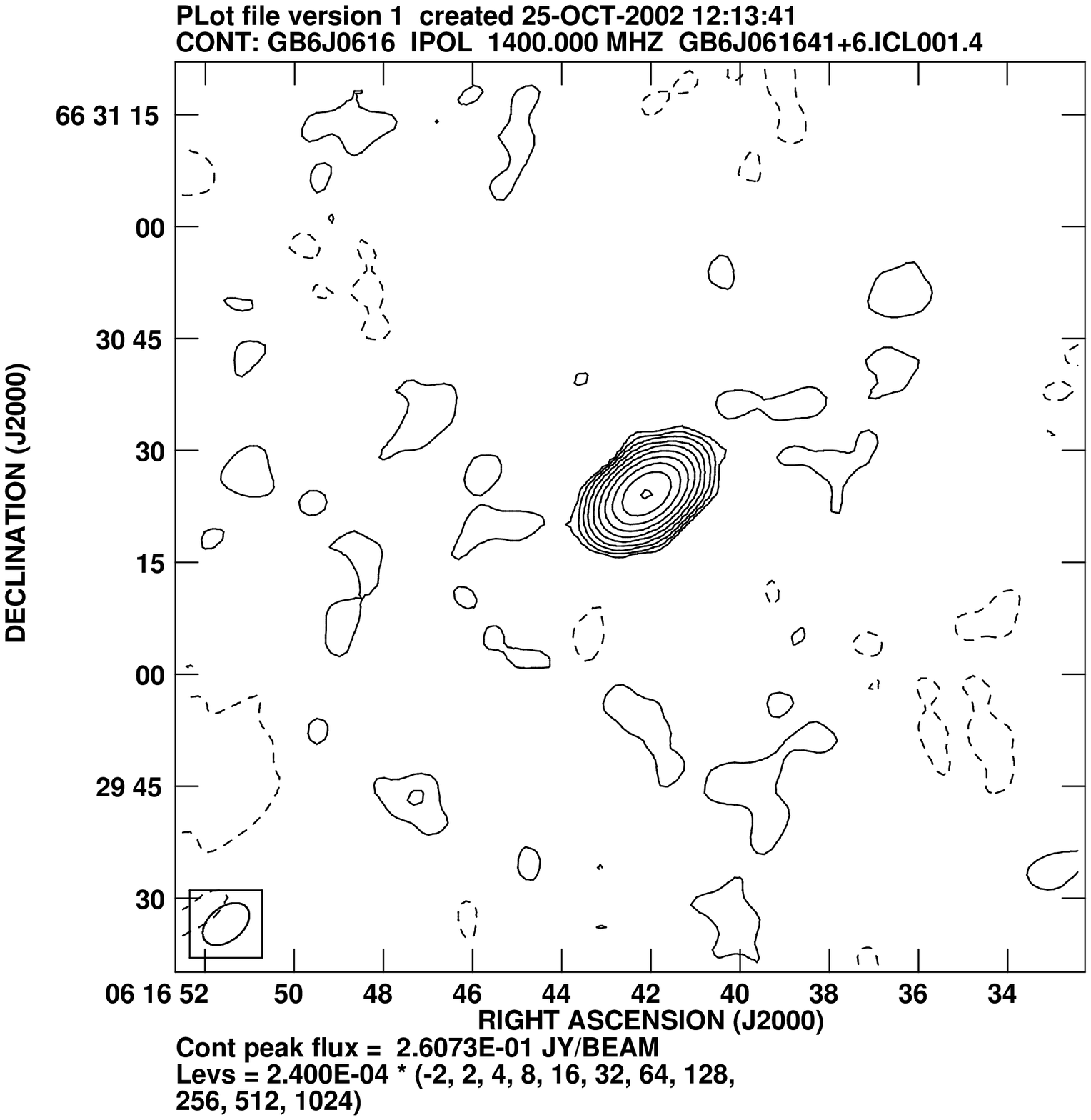,height=7cm,width=7cm} \\
 GB6J070932+501056 & GB6J073728+594106 \\
\psfig{figure=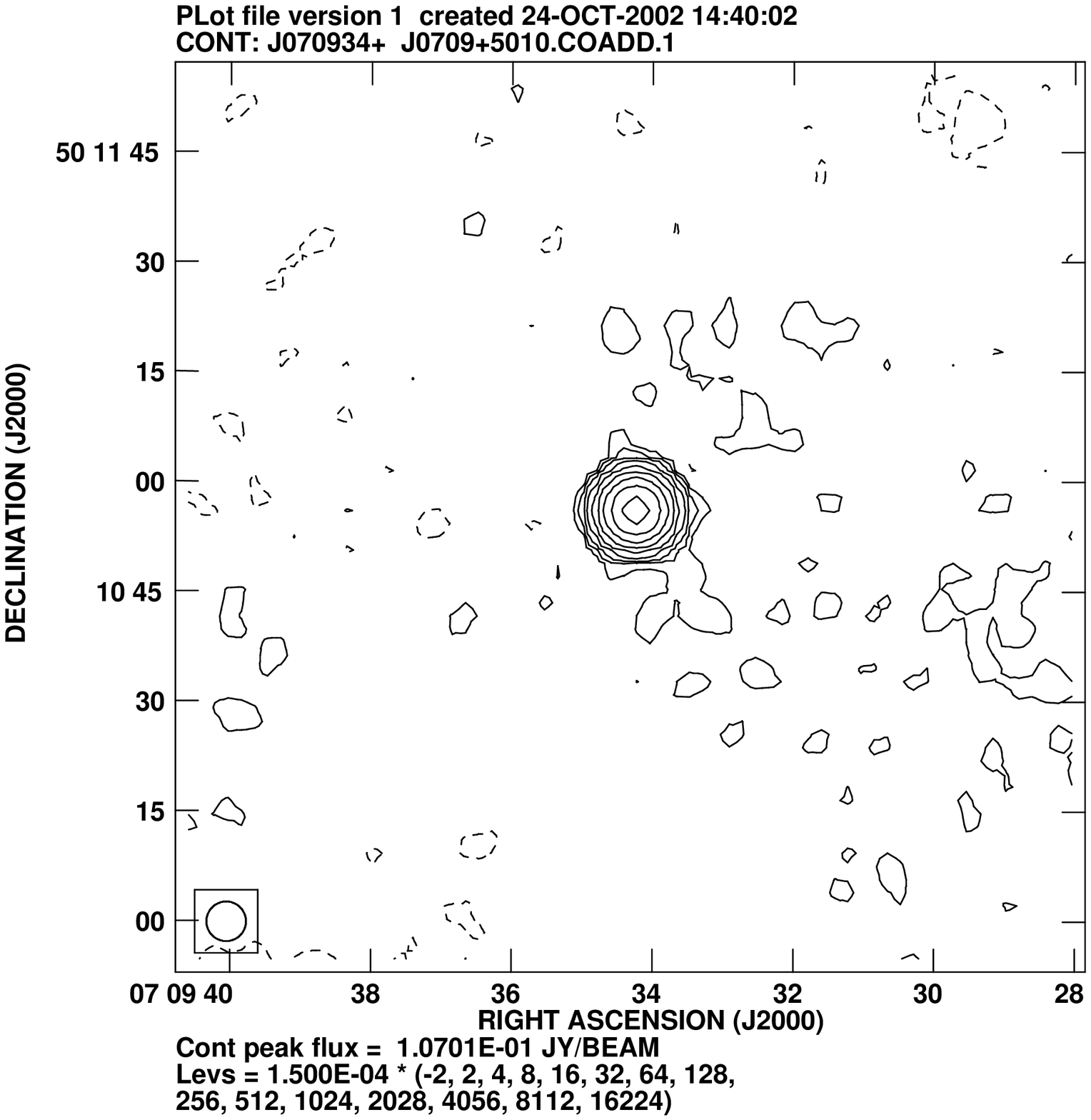,height=7cm,width=7cm} &
\psfig{figure=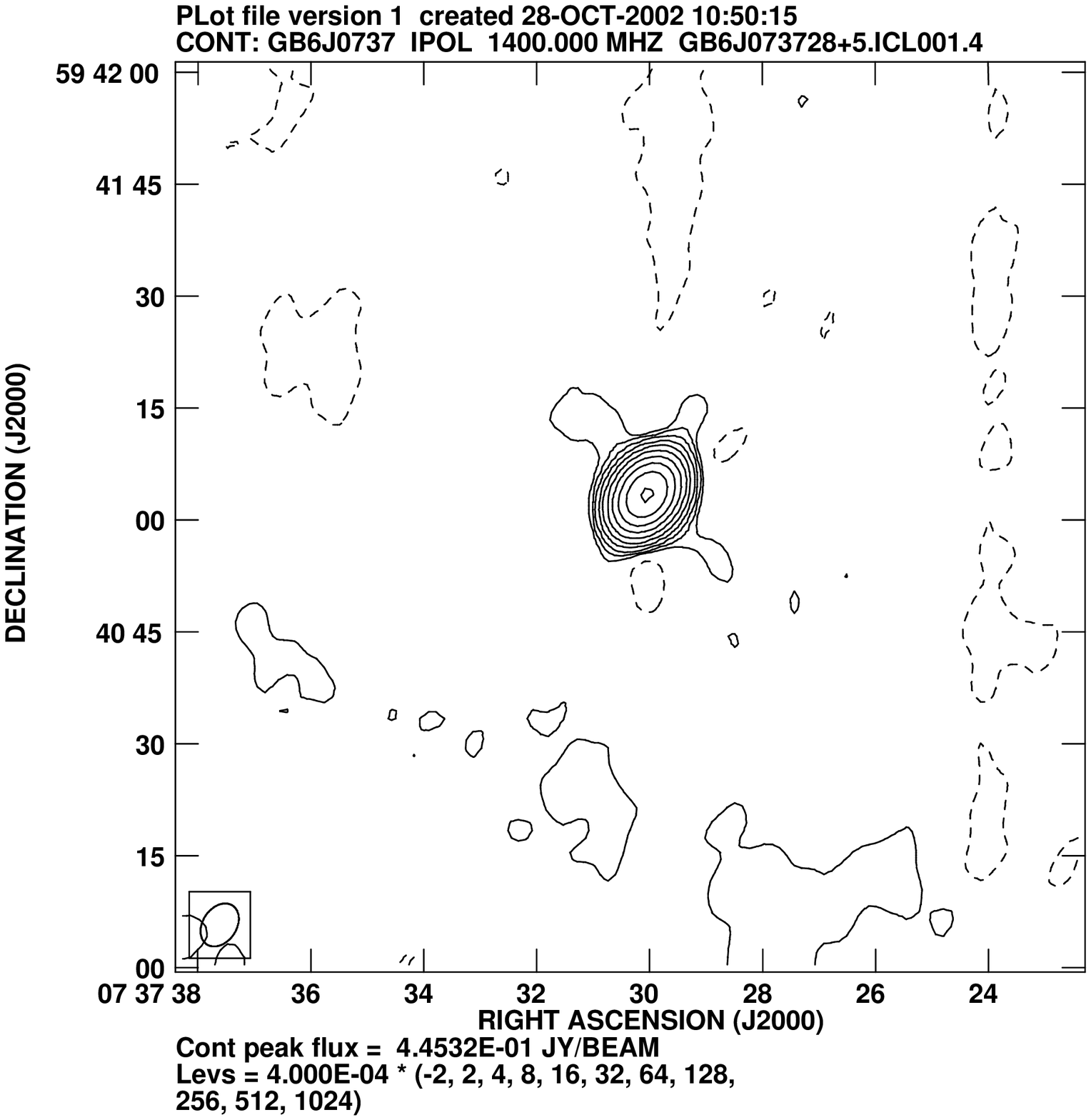,height=7cm,width=7cm} \\
 GB6J090615+463633 & GB6J094319+361447 \\
\psfig{figure=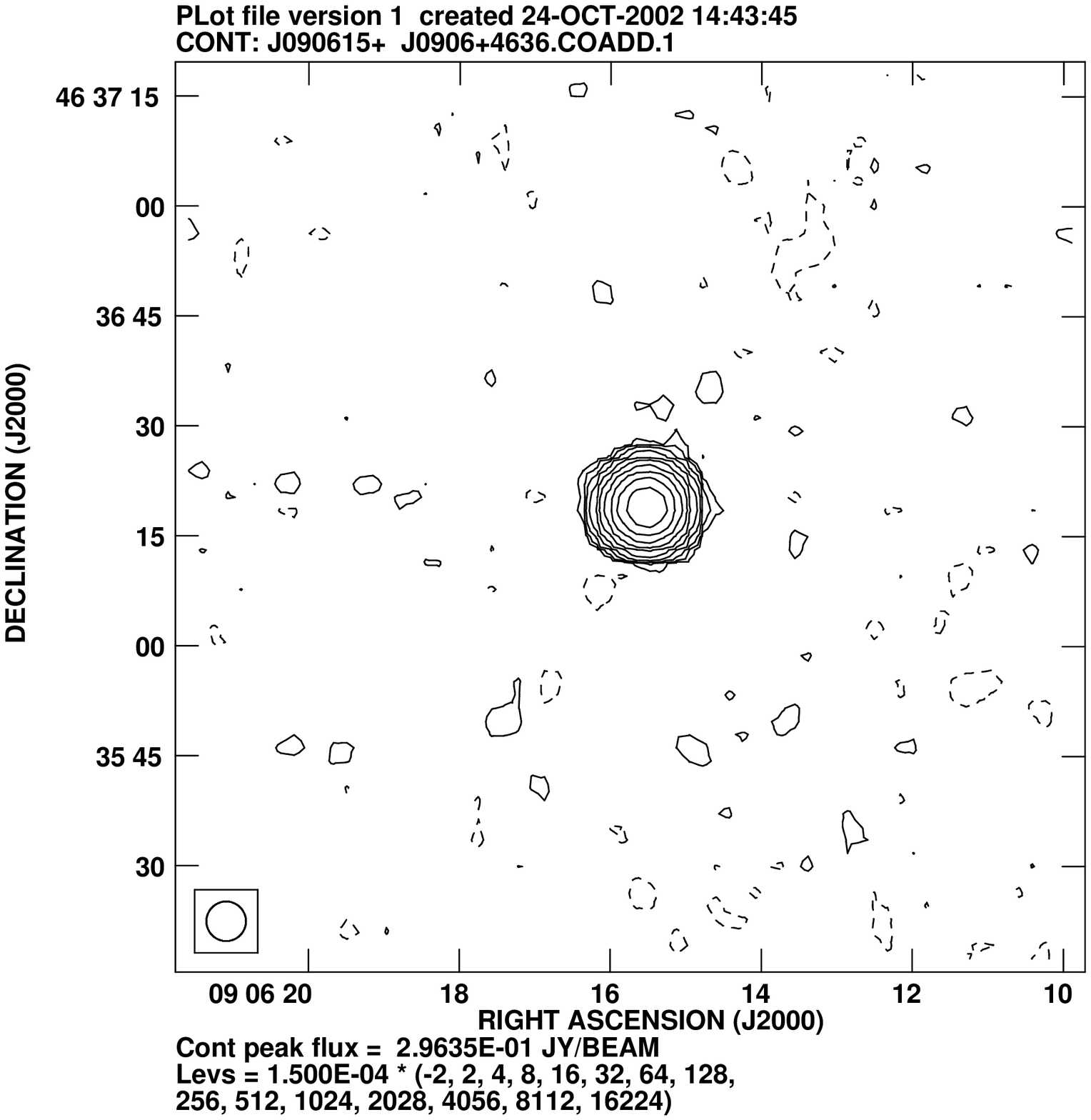,height=7cm,width=7cm} &
\psfig{figure=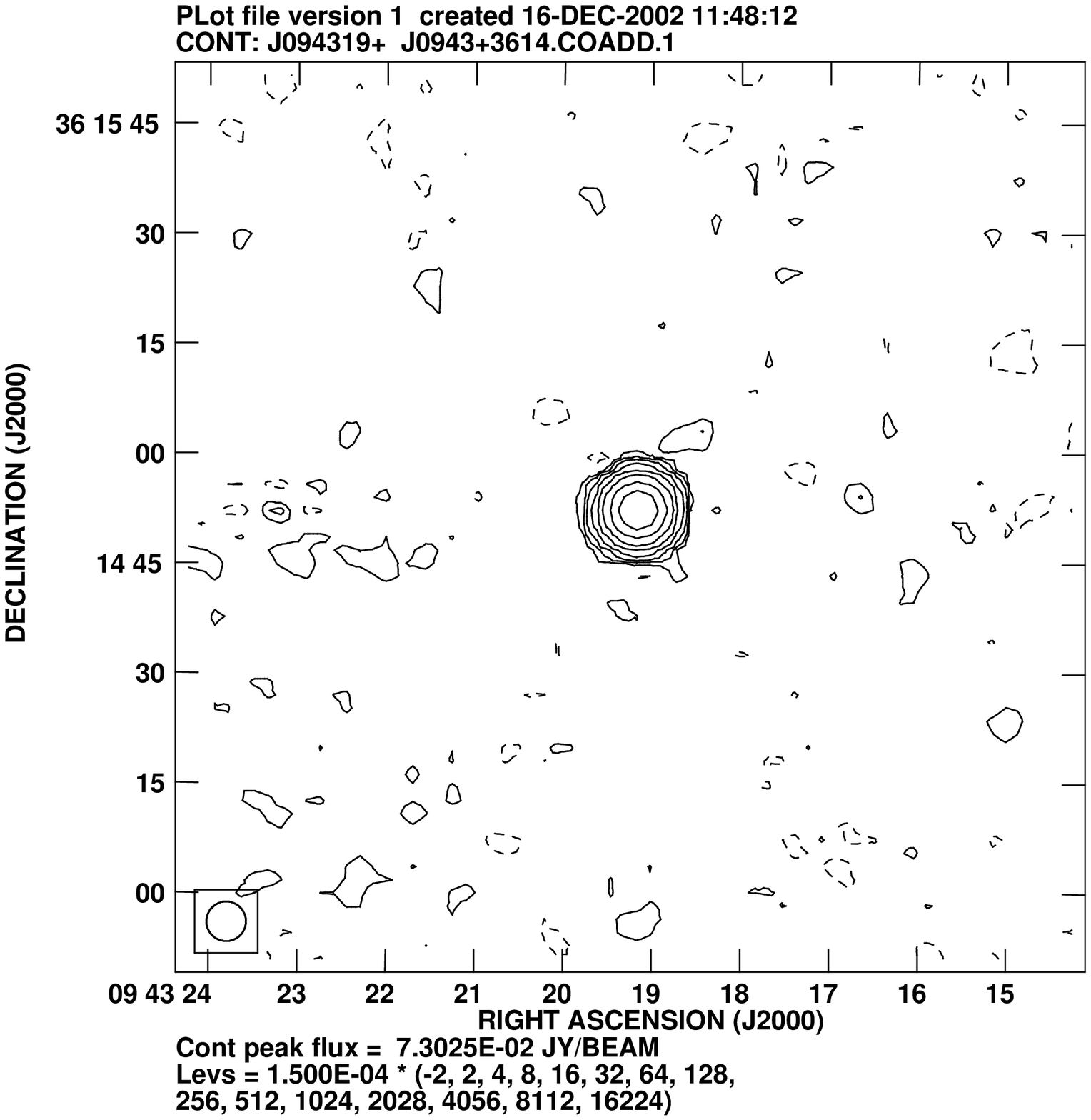,height=7cm,width=7cm} \\
\end{tabular}
\caption{The VLA B-array maps at 1.4 GHz of the blazar candidates with 
$\alpha_{RX}>$0.75 and P$_{5 GHz}$ between 10$^{23}$ and 10$^{25}$ W Hz$^{-1}$.
Given the redshift of these sources, the resolution of these maps 
corresponds to a linear scale between 1.4-14 kpc.}
\label{map}
\newpage
\end{figure}

\begin{figure}
\begin{tabular}{c c}
 GB6J114850+592459 & GB6J120304+603130 \\
\psfig{figure=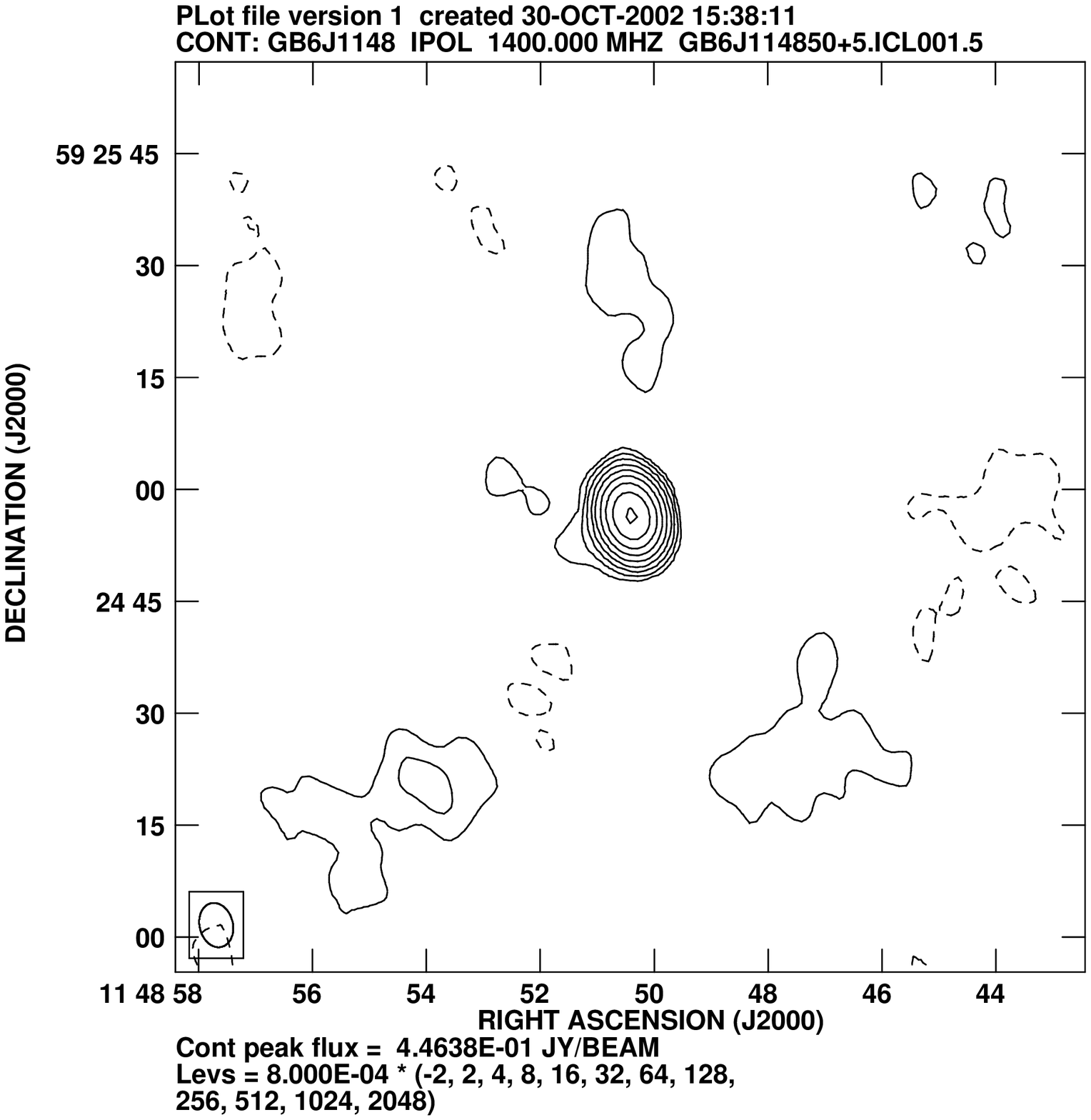,height=7cm,width=7cm} &
\psfig{figure=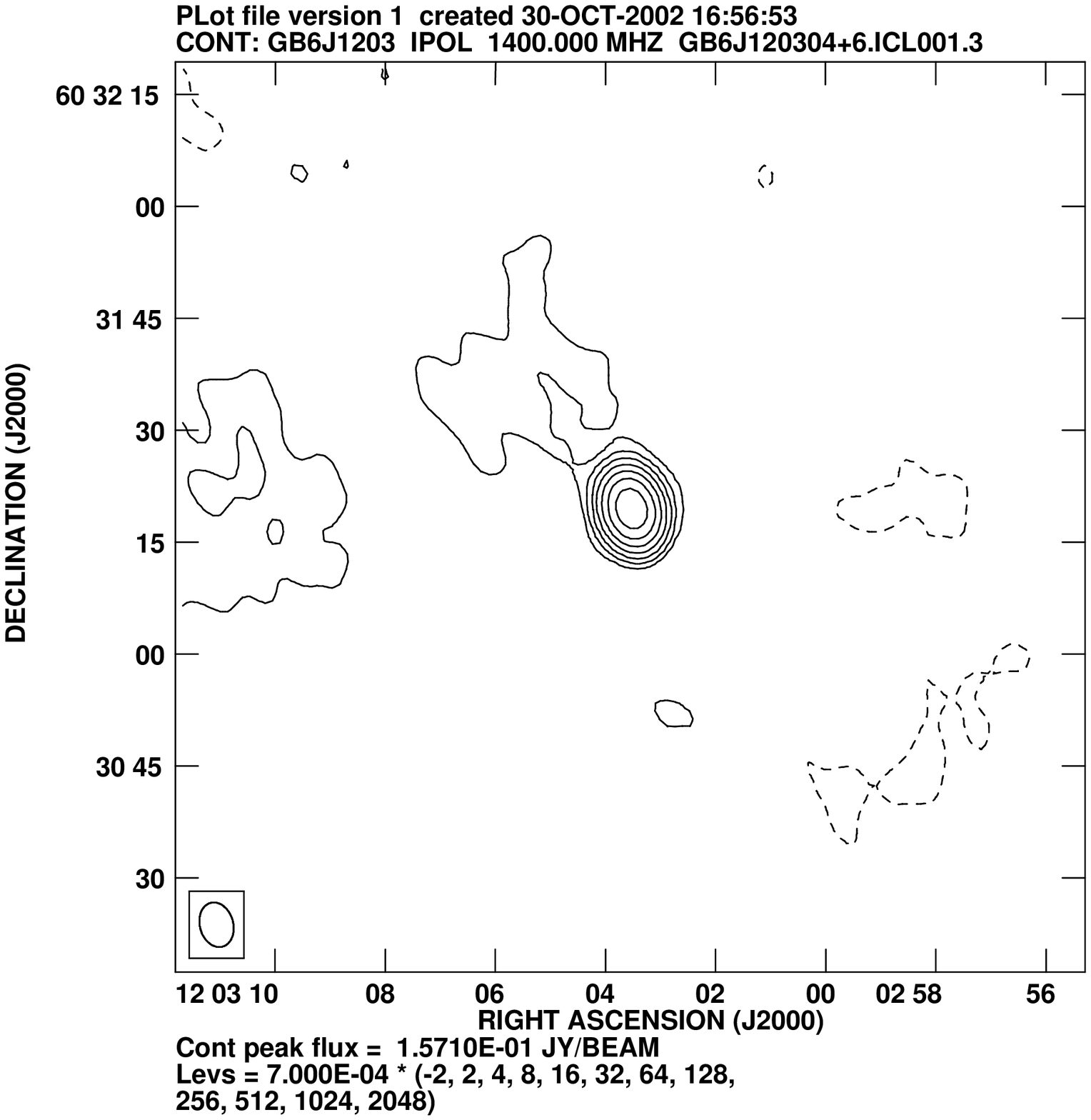,height=7cm,width=7cm} \\
 GB6J121331+504446 & GB6J123012+470031 \\
\psfig{figure=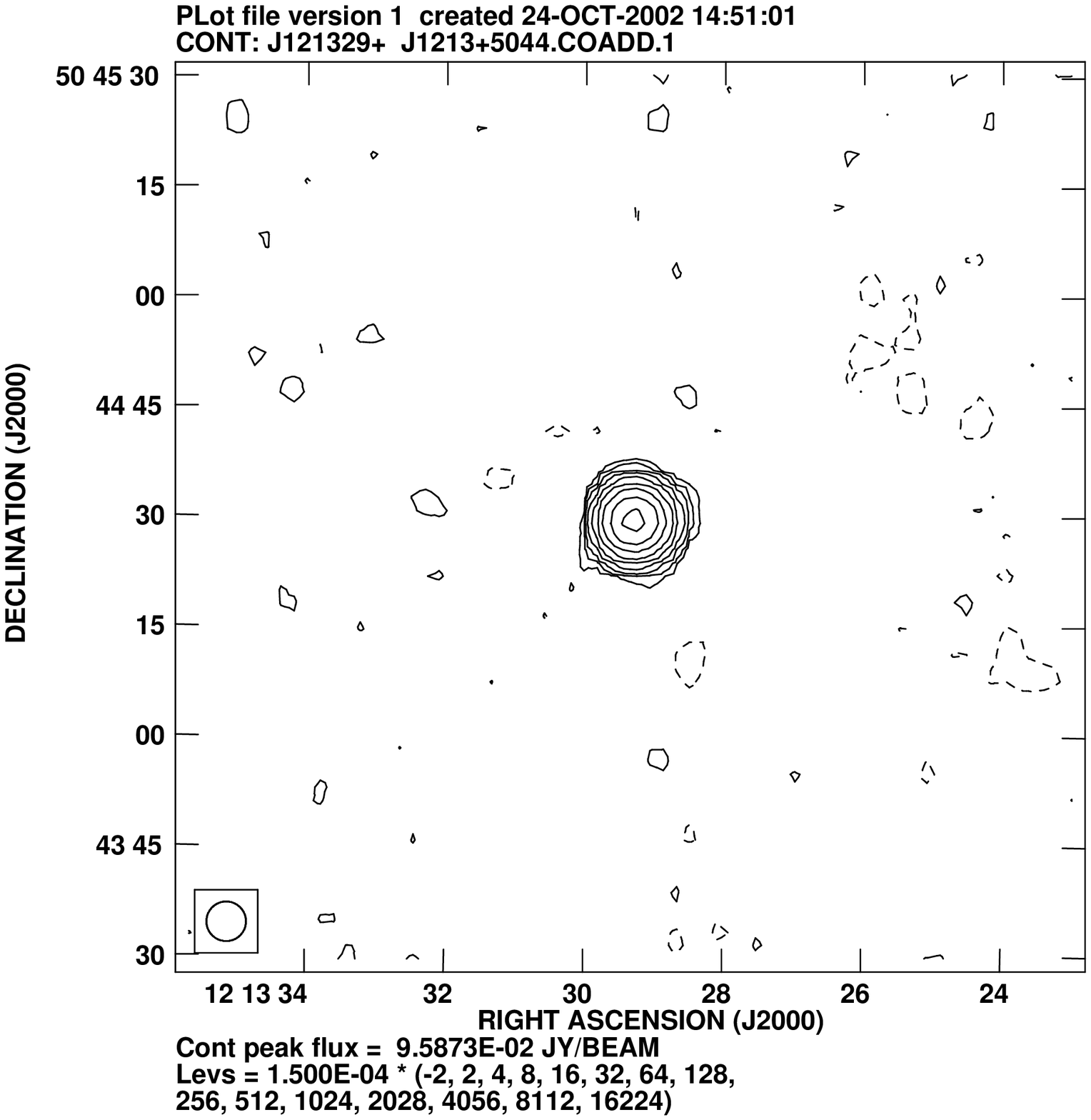,height=7cm,width=7cm} &
\psfig{figure=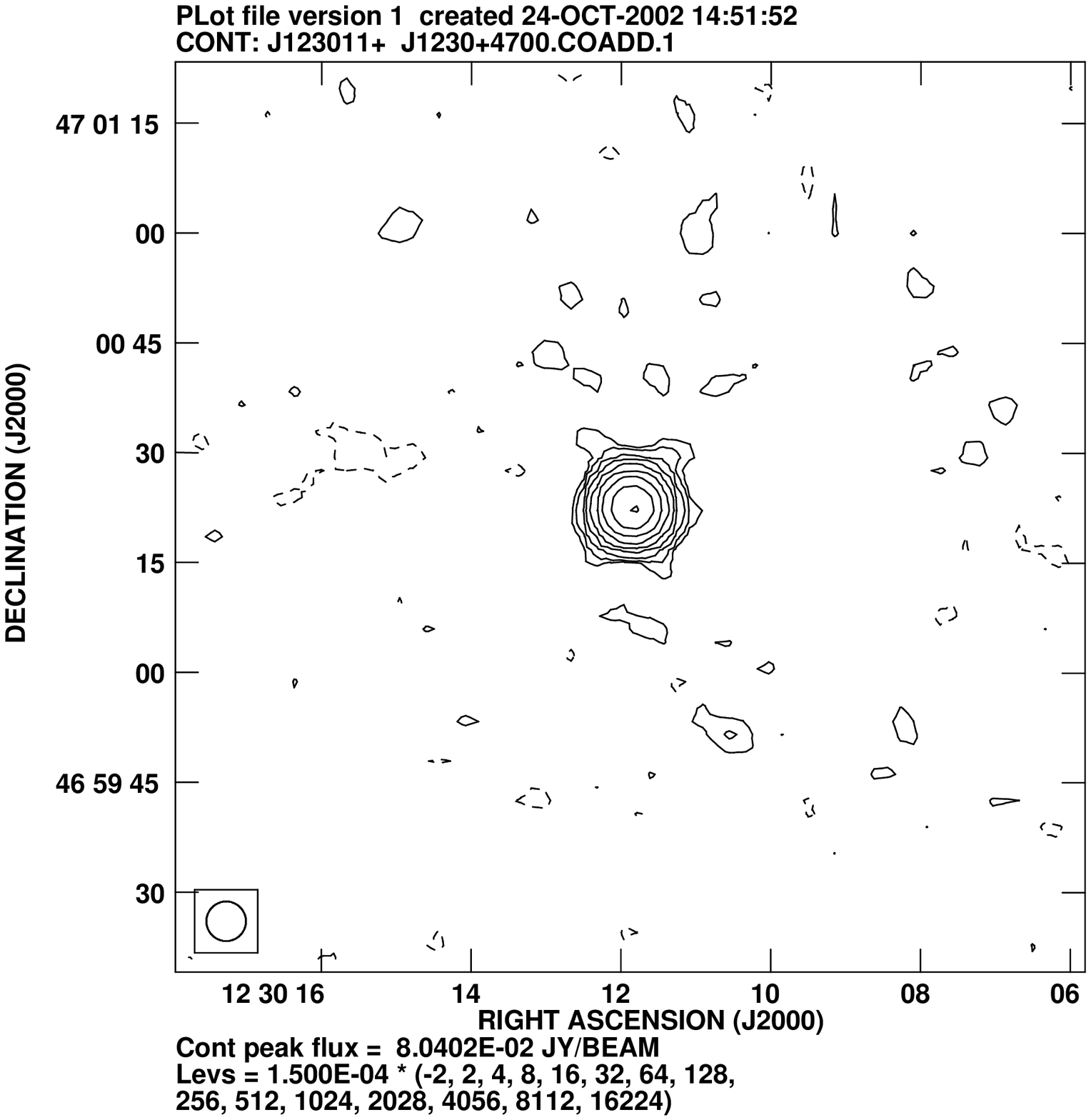,height=7cm,width=7cm} \\
 GB6J124732+672322 & GB6J125614+565220 \\
\psfig{figure=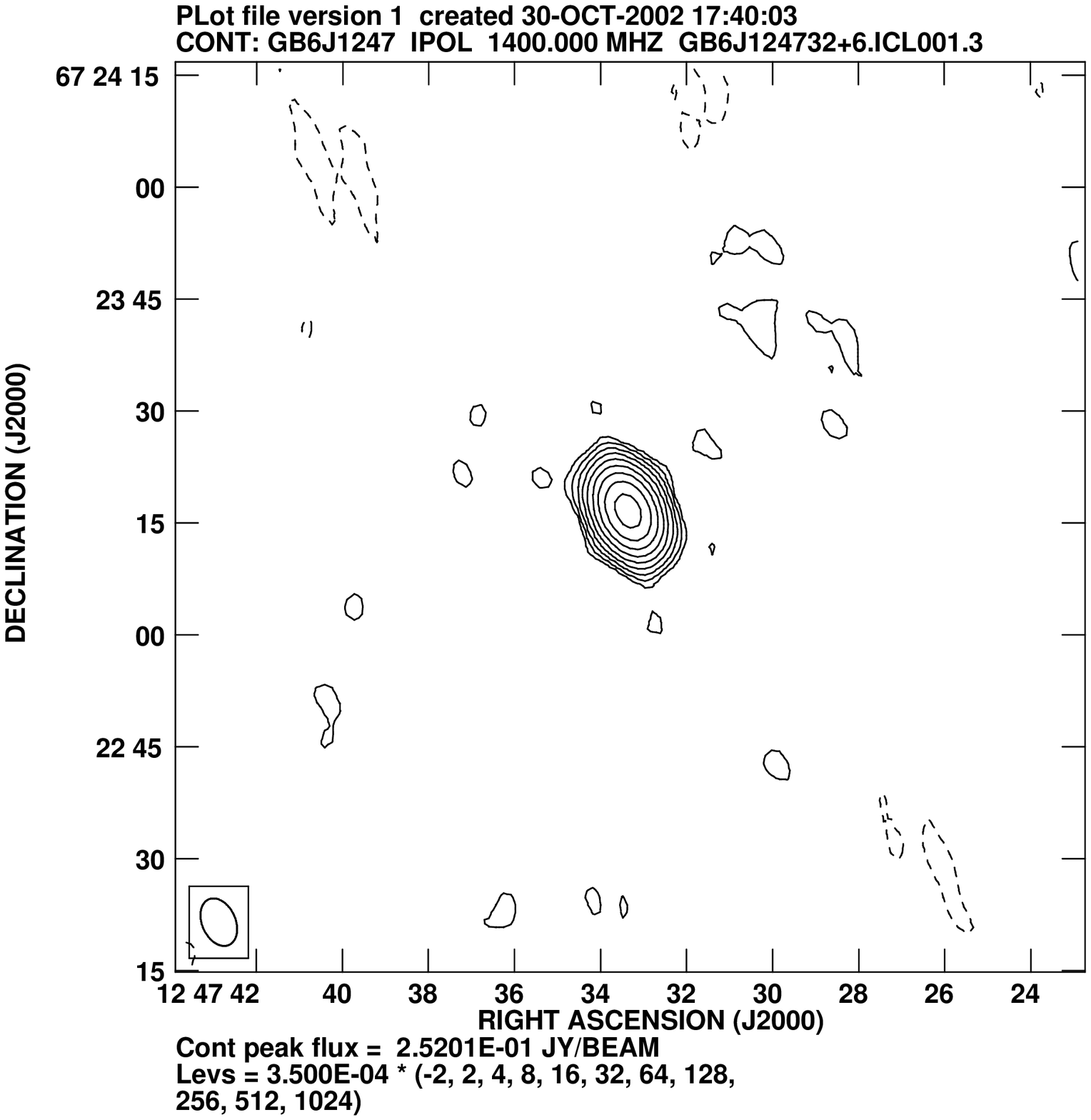,height=7cm,width=7cm} &
\psfig{figure=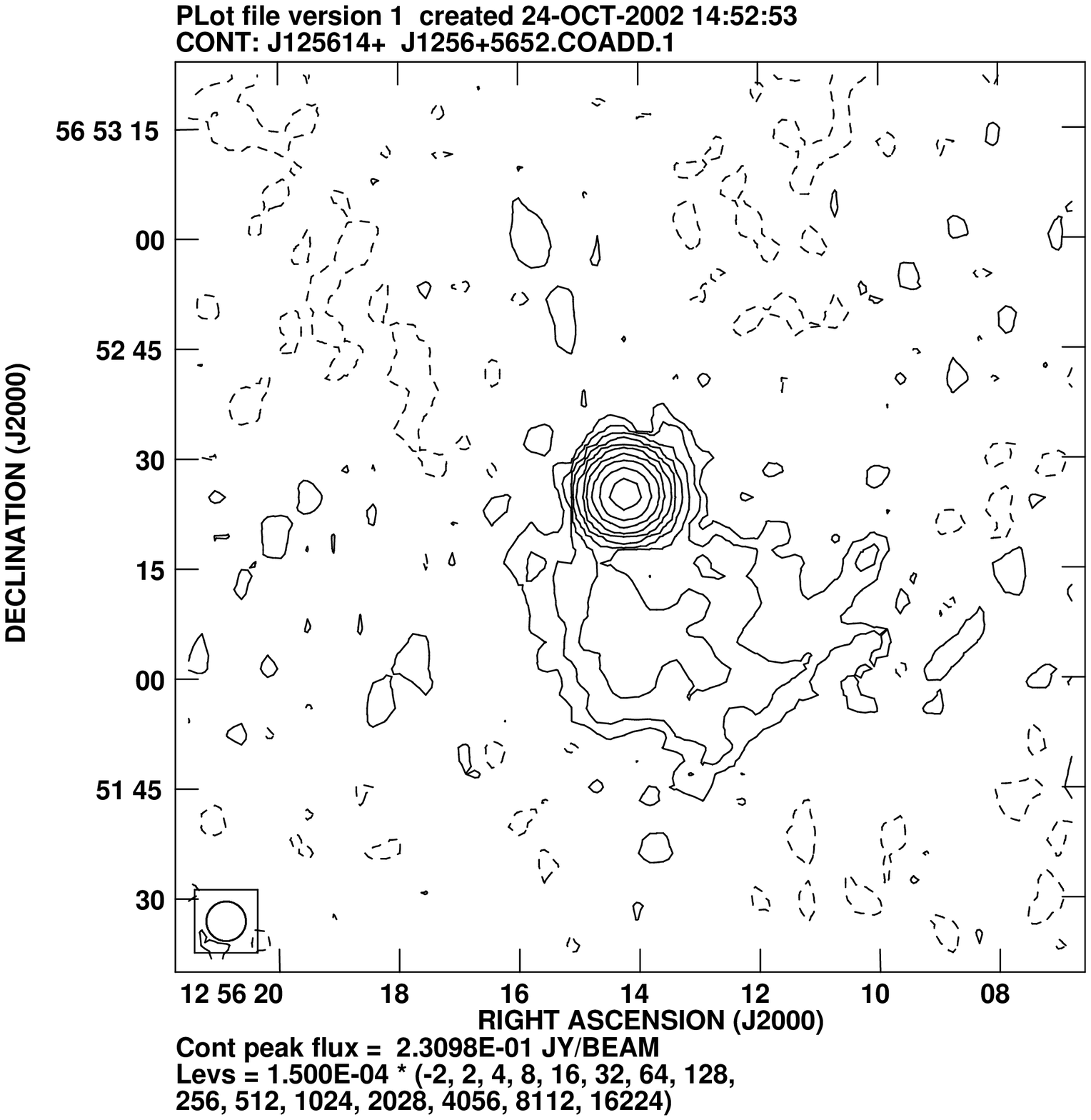,height=7cm,width=7cm} \\
\end{tabular}
\contcaption{}
\newpage
\end{figure}

\begin{figure}
\begin{tabular}{c c}

 GB6J131739+411538 & GB6J134035+444801 \\
\psfig{figure=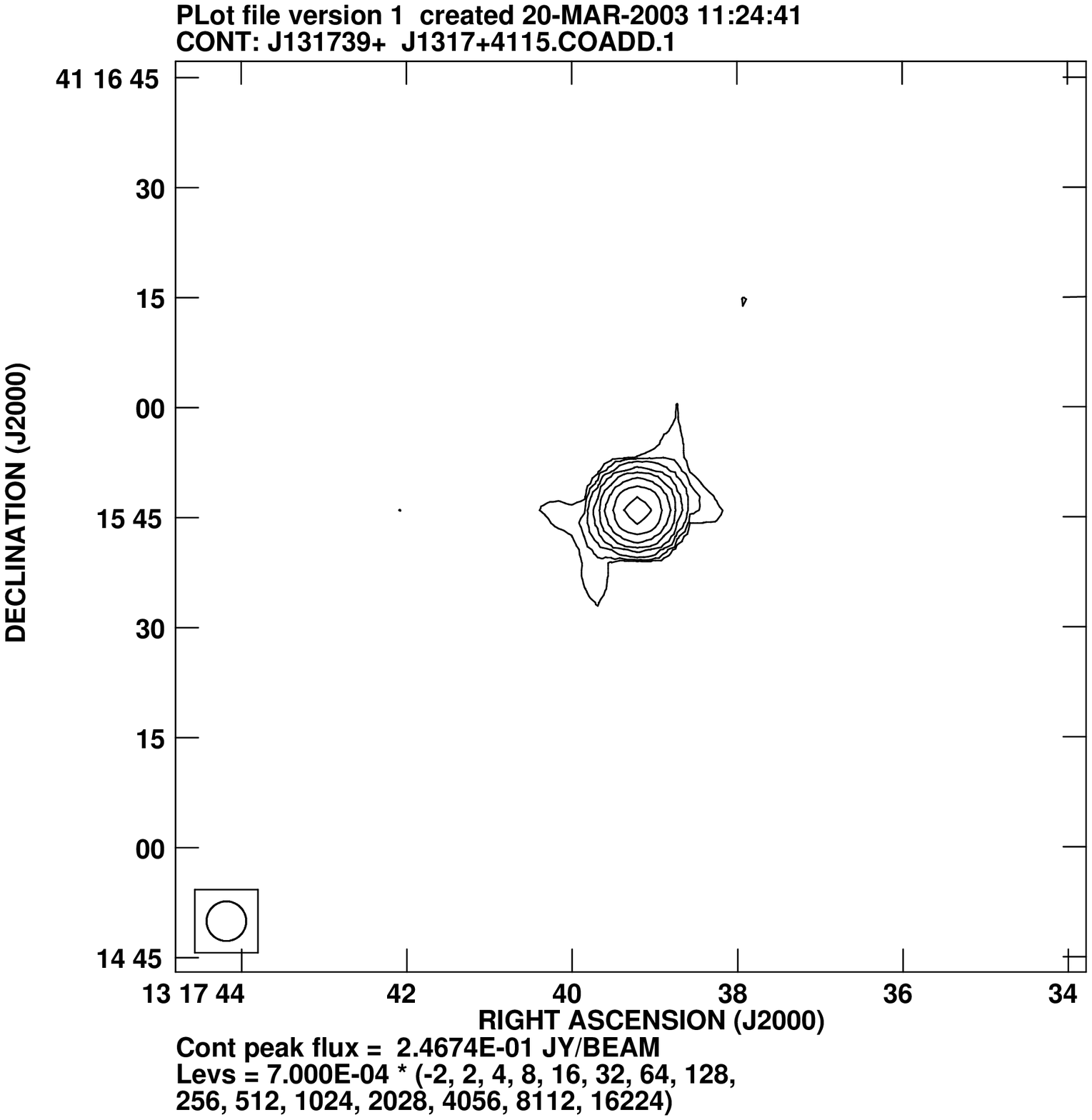,height=7cm,width=7cm} &
\psfig{figure=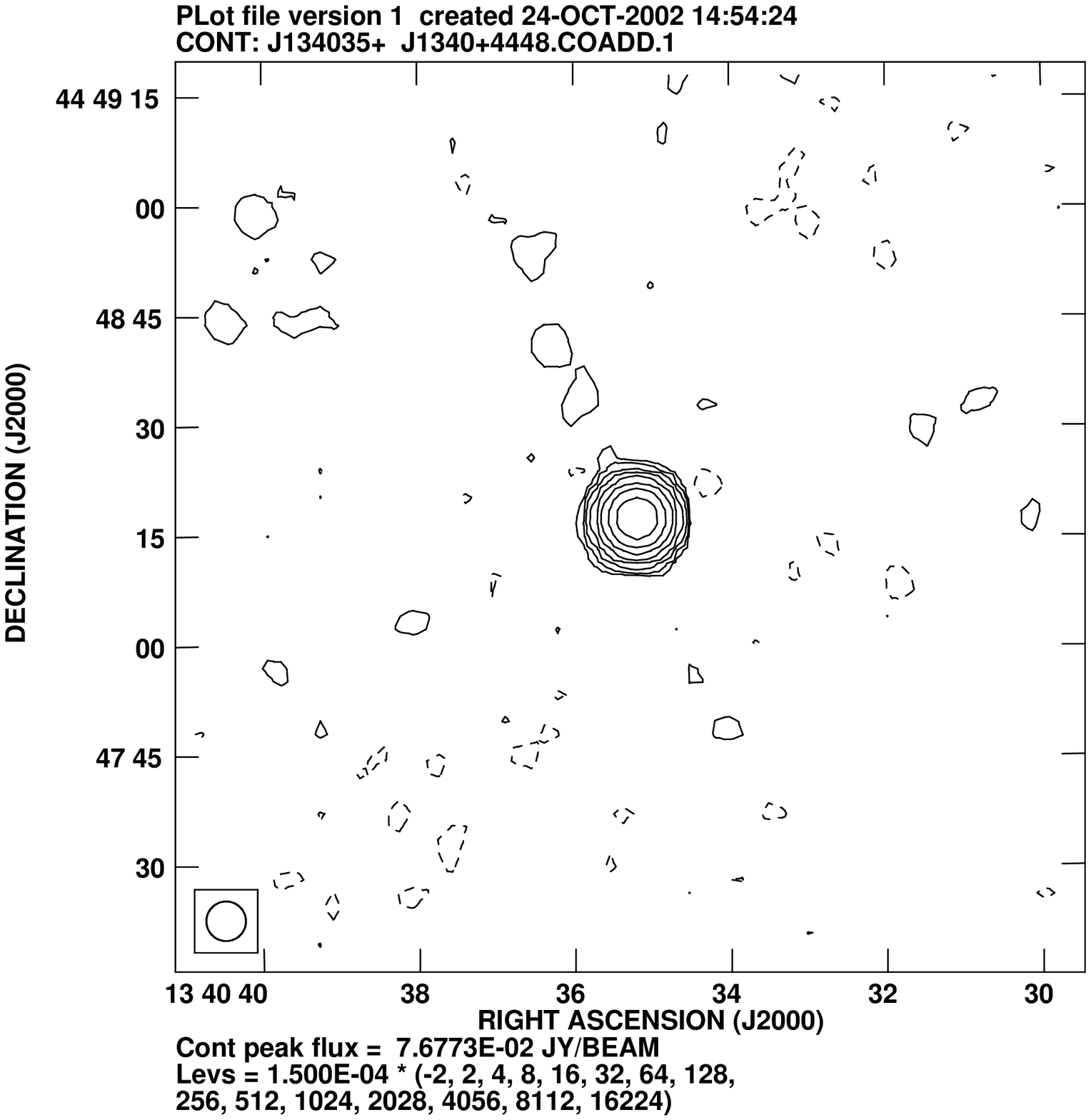,height=7cm,width=7cm} \\
 GB6J134444+555322 & GB6J143239+361823 \\
\psfig{figure=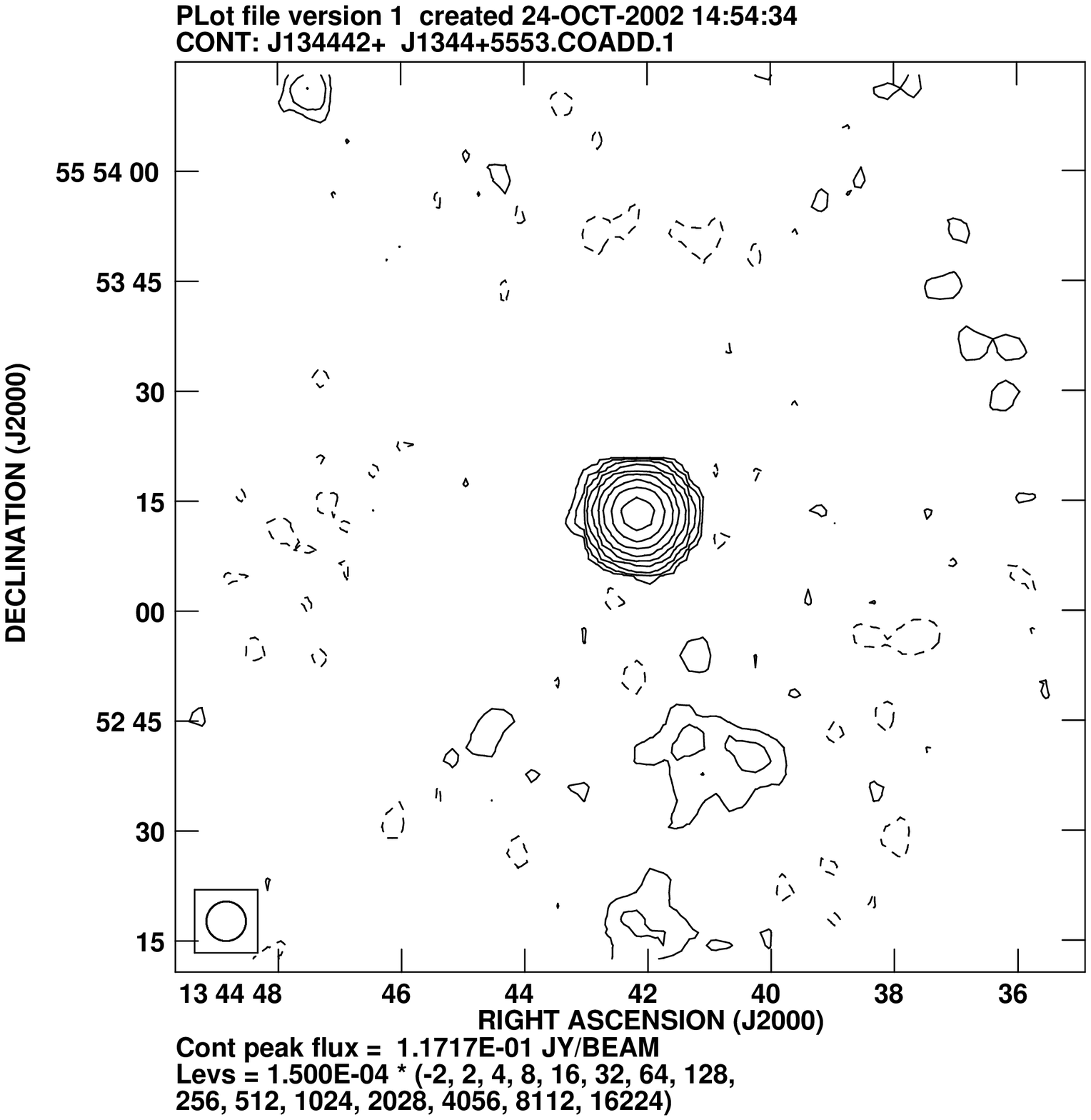,height=7cm,width=7cm} &
\psfig{figure=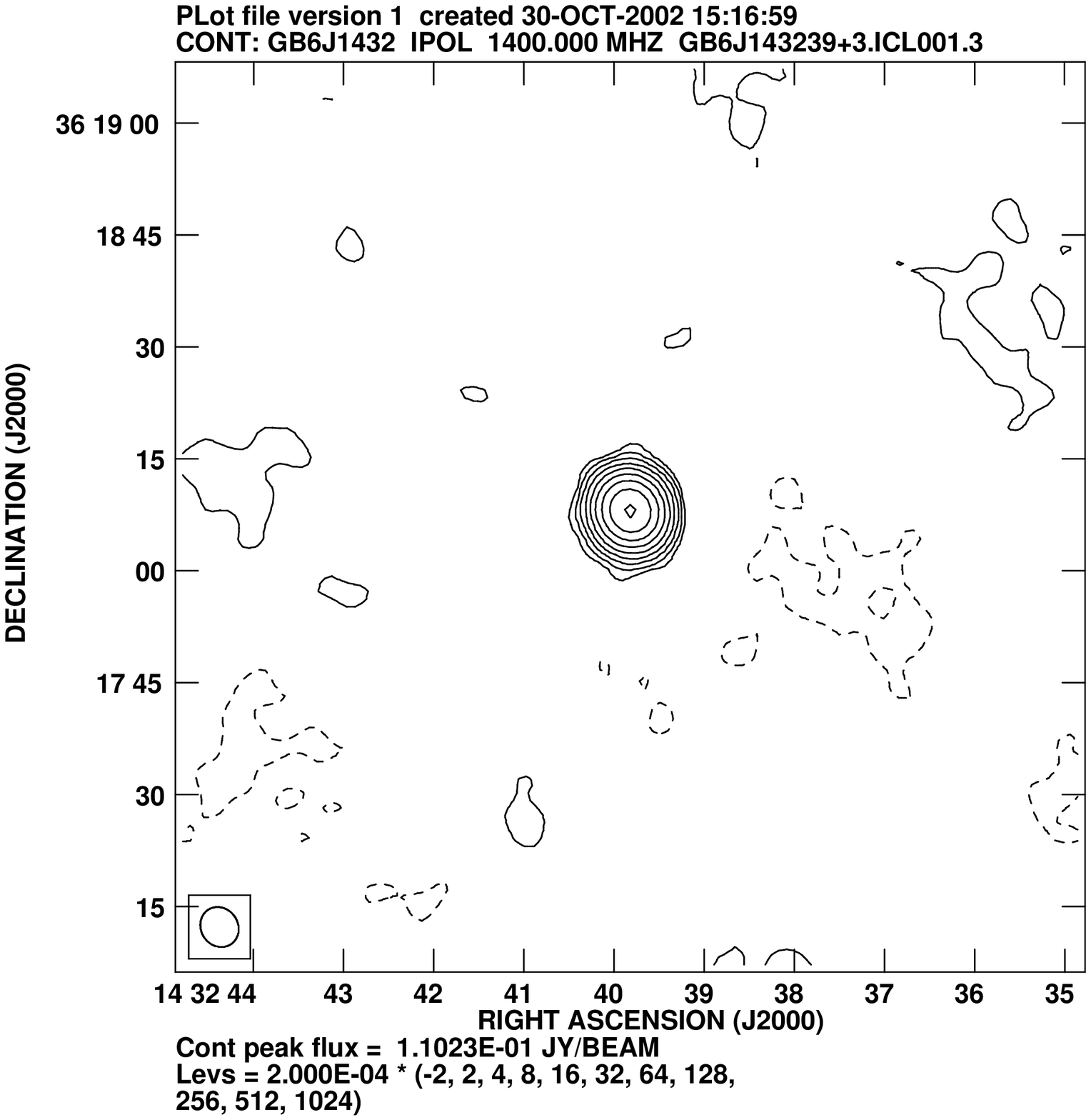,height=7cm,width=7cm} \\
 GB6J153900+353053 & GB6J155901+592437 \\
\psfig{figure=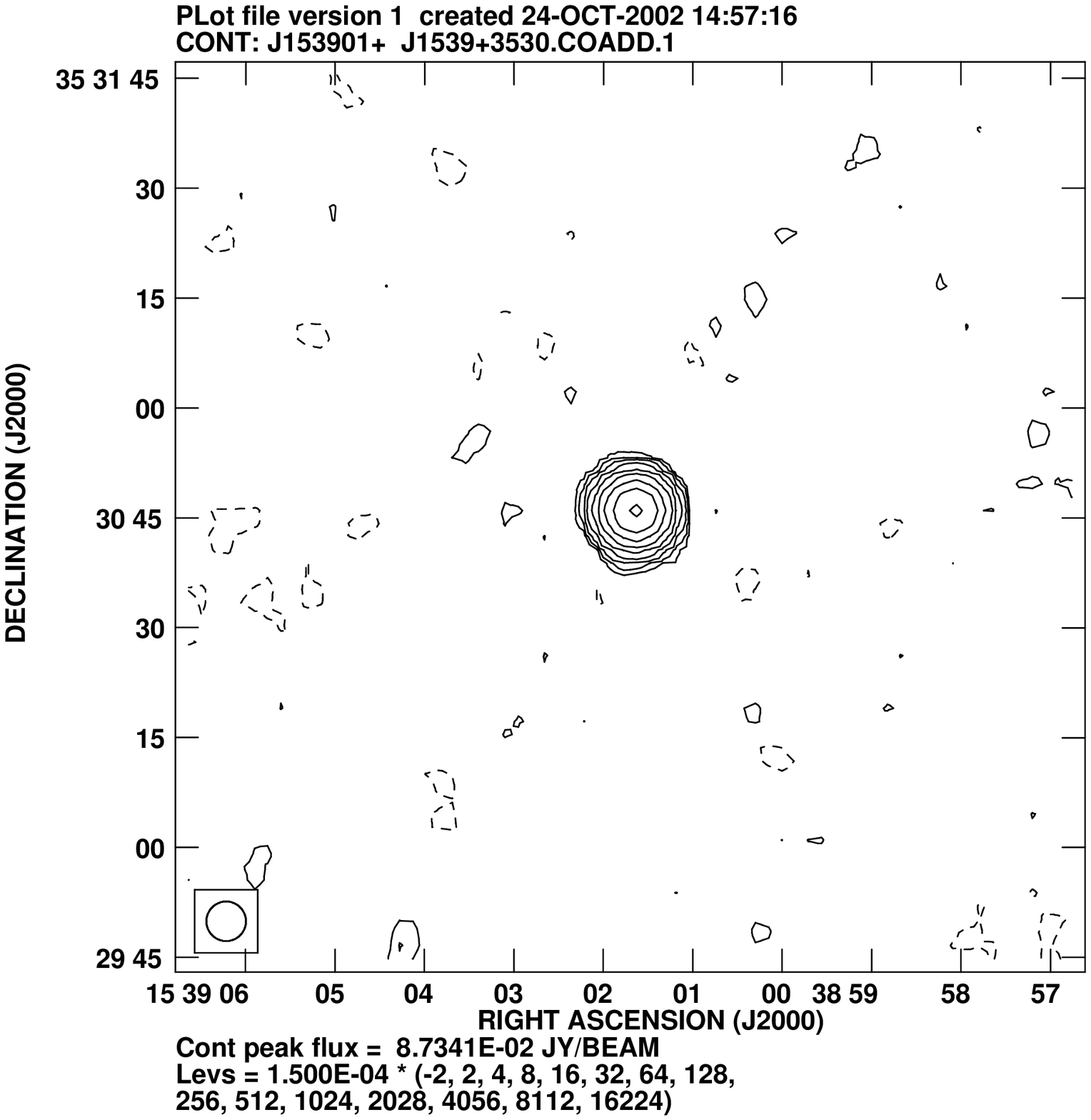,height=7cm,width=7cm} &
\psfig{figure=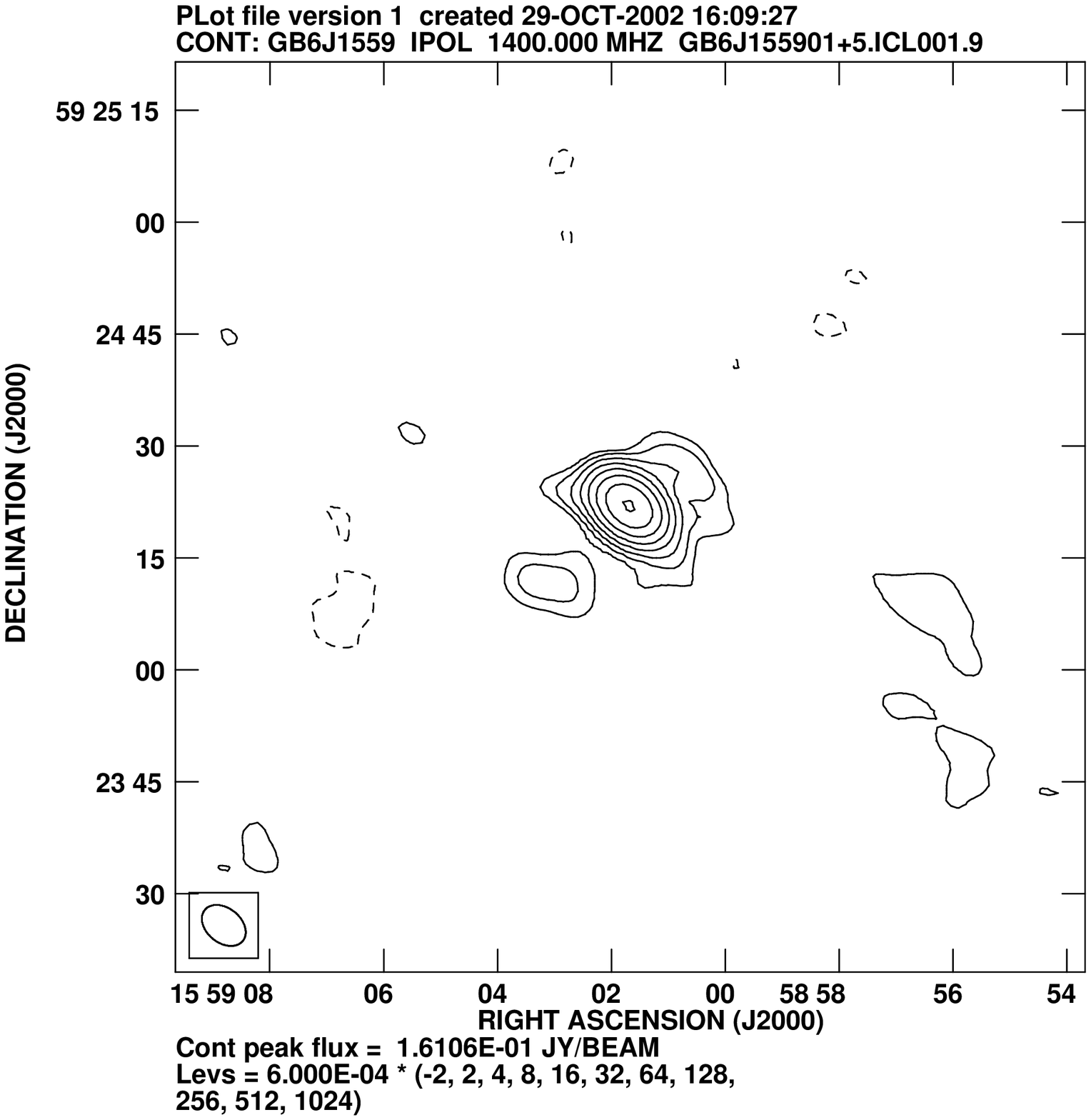,height=7cm,width=7cm} \\

\end{tabular}
\contcaption{}
\newpage
\end{figure}

\begin{figure}
\begin{tabular}{c c}
 GB6J162509+405345 & GB6J164734+494954 \\
\psfig{figure=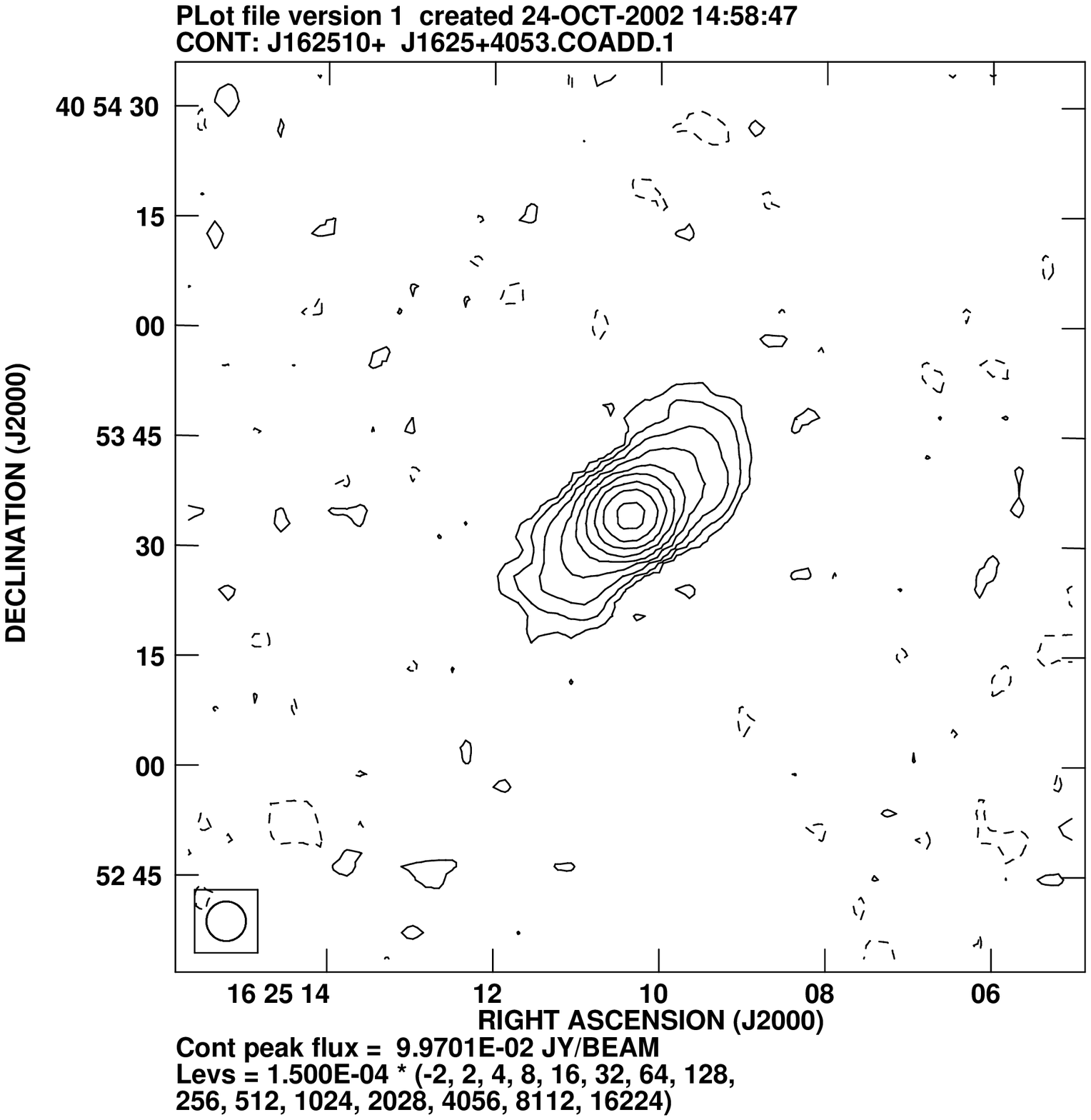,height=7cm,width=7cm} &
\psfig{figure=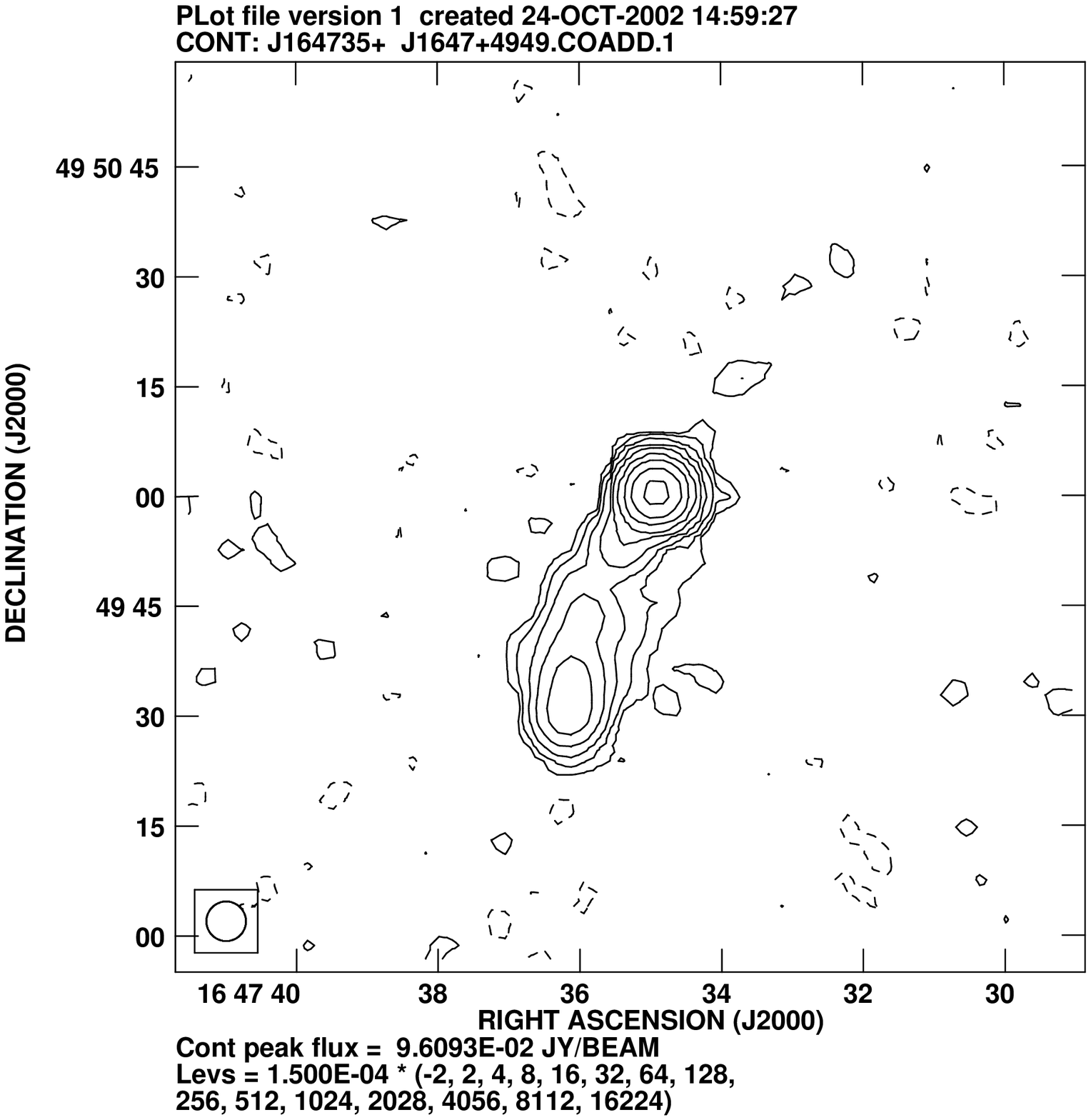,height=7cm,width=7cm} \\
 GB6J171914+485839 & GB6J174455+554220 \\
\psfig{figure=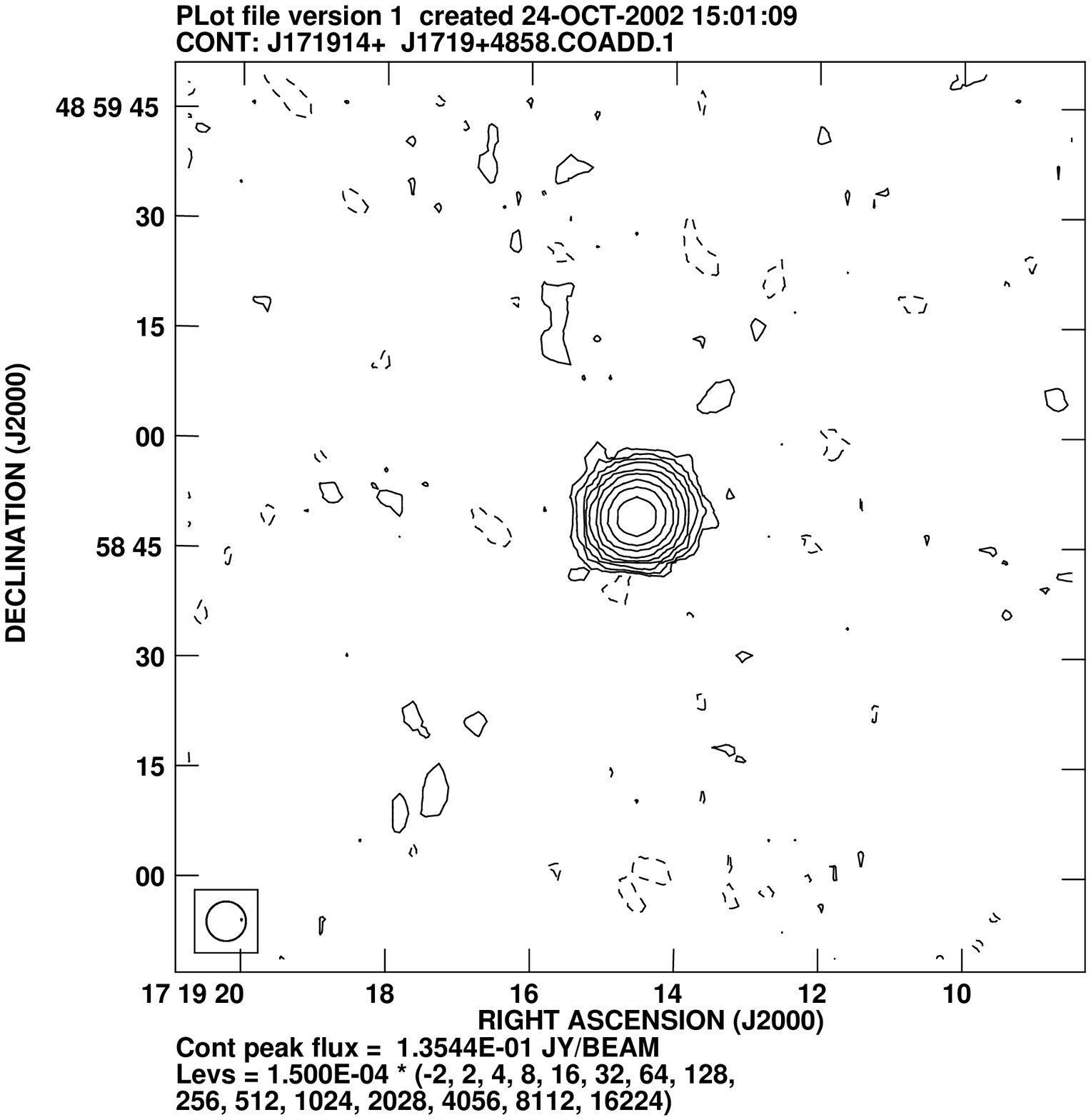,height=7cm,width=7cm} &
\psfig{figure=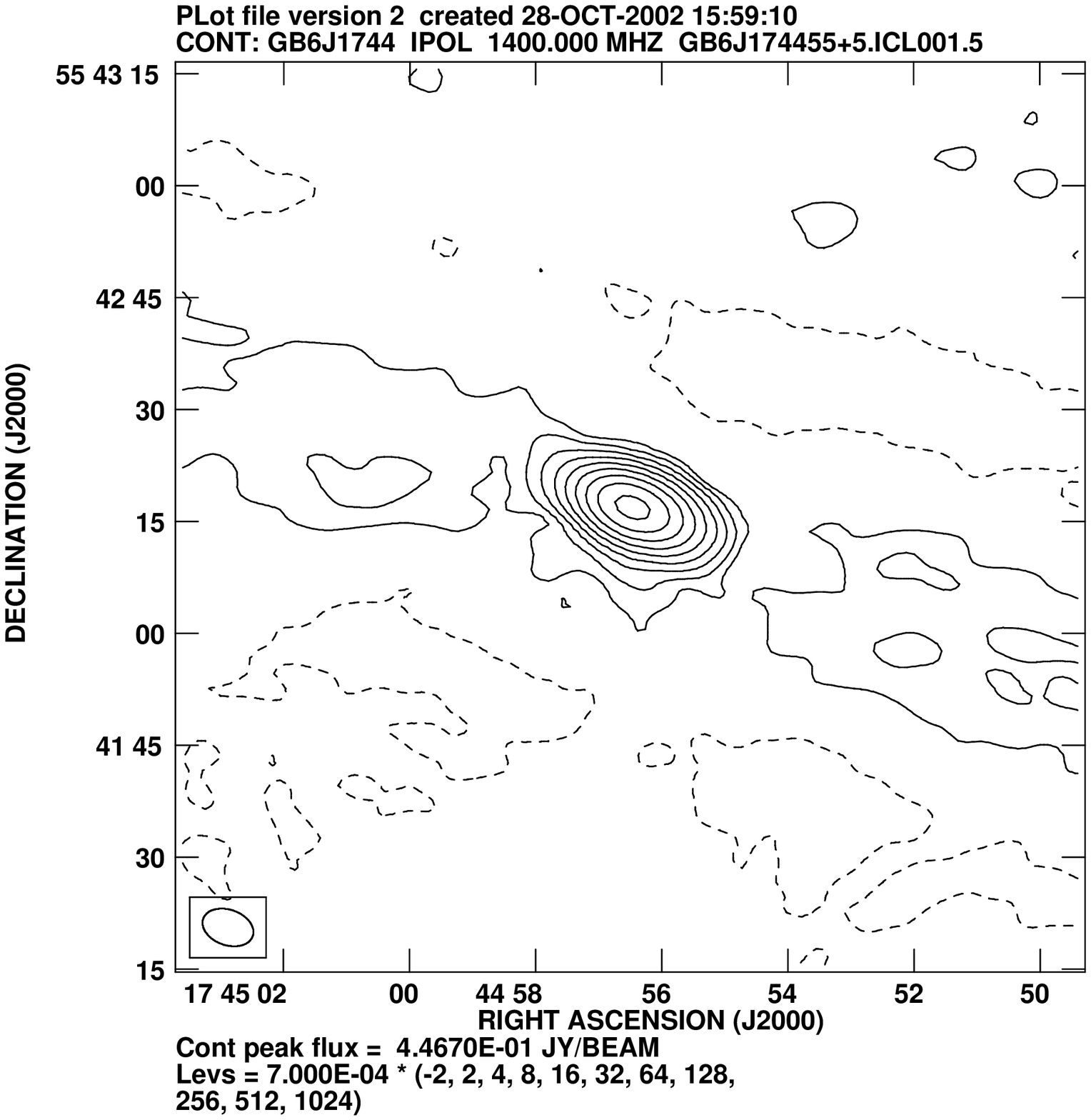,height=7cm,width=7cm} \\
 GB6J175546+623652 & GB6J175728+552309 \\
\psfig{figure=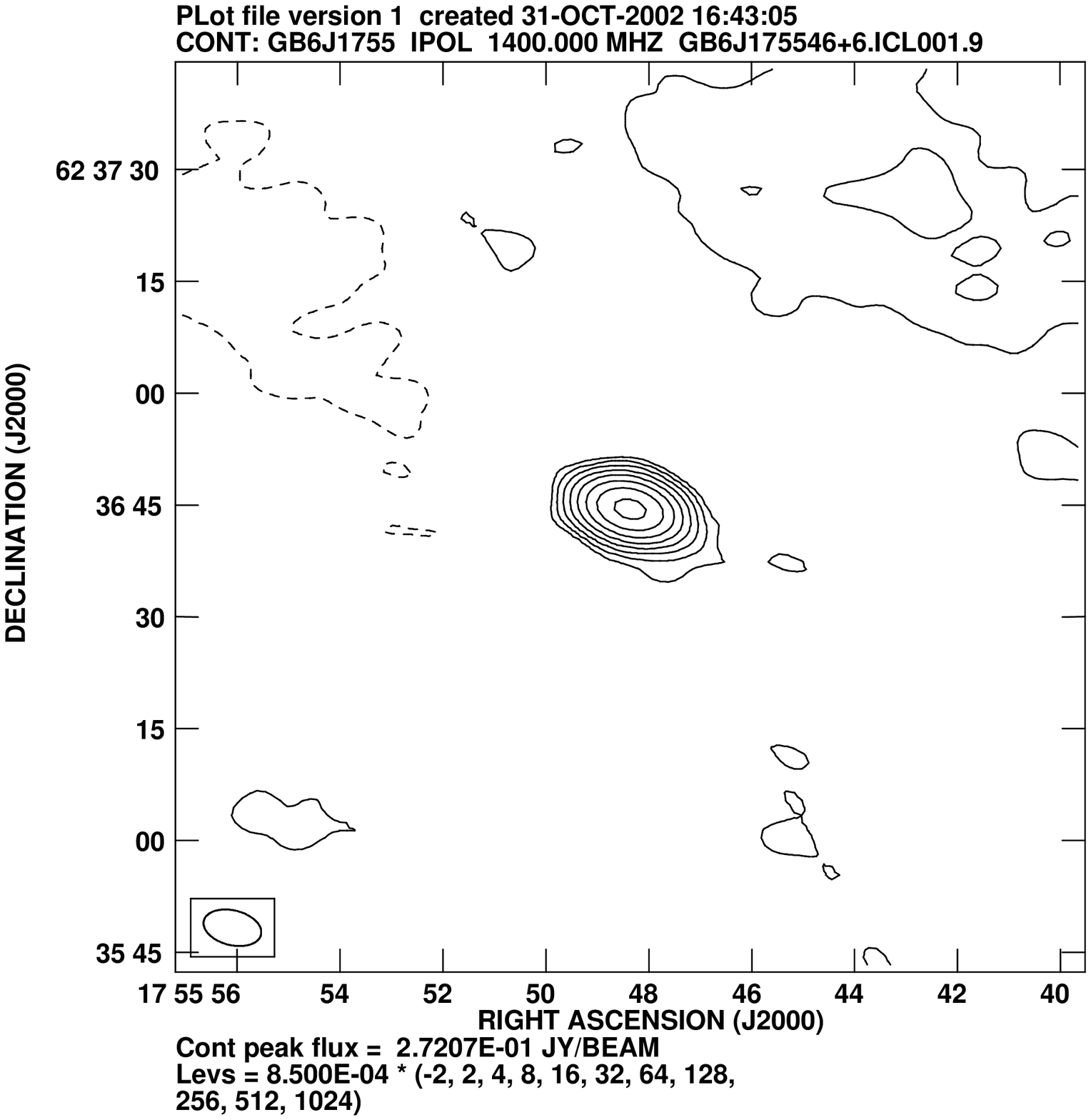,height=7cm,width=7cm} &
\psfig{figure=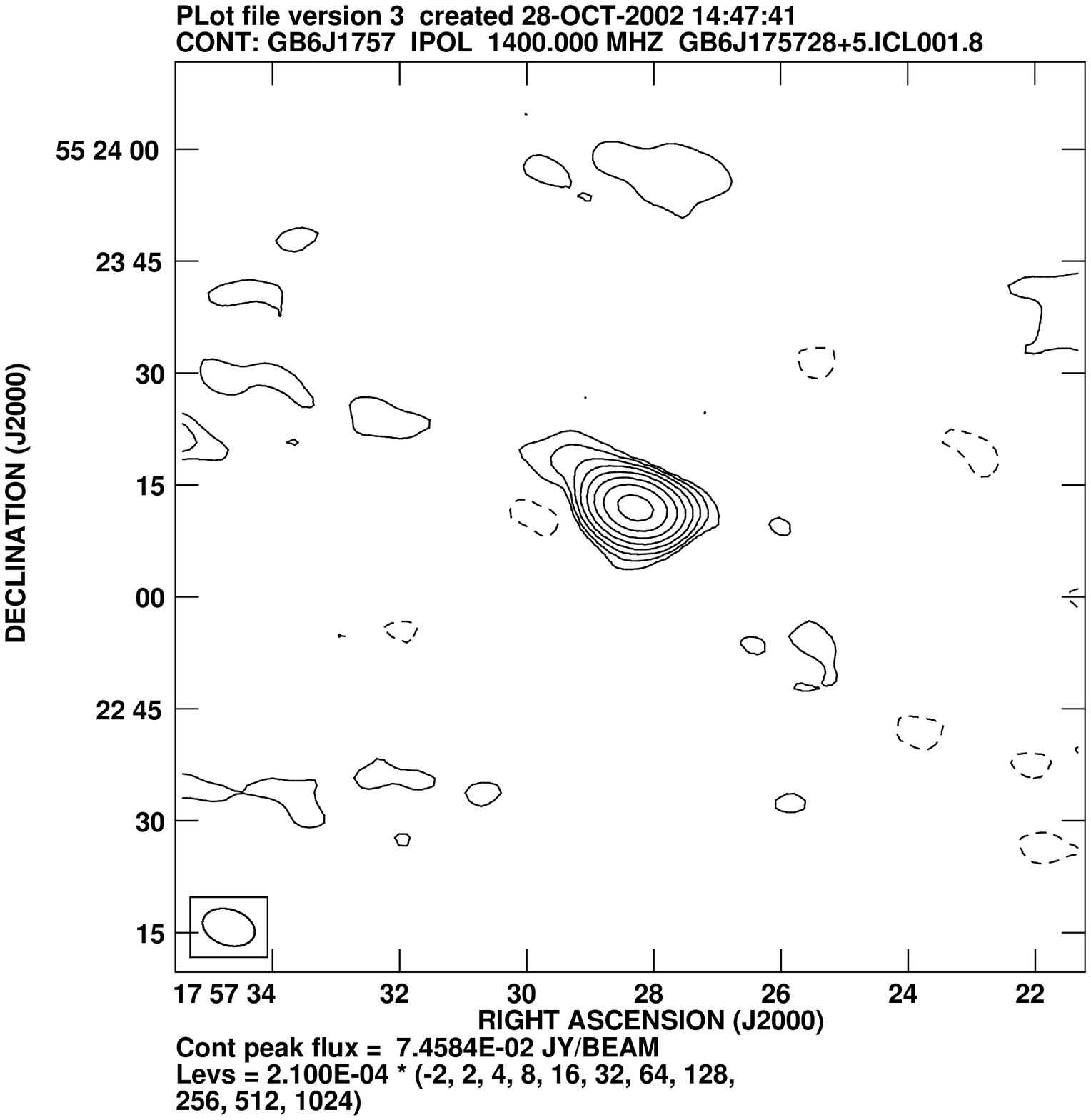,height=7cm,width=7cm} \\
\end{tabular}
\contcaption{}
\newpage
\end{figure}

\end{onecolumn}

\begin{table*}
\caption{VLA data}
\begin{tabular}{l r r r l l l}
name & core flux & total flux & R & data (core) & data (total) & comments \\ 
GB6J002538+400830 &      17.00 &      36.90 &          0.77 & may11 & NVSS &   \\
GB6J012626+395406 &     $<$1.4 &     175.00 &       $<$0.01 & u.l. & NVSS & double \\
GB6J013631+390623 &      58.80 &      58.80 &      $>$588 & may11 & B &   \\
GB6J014156+392325 &      94.90 &      97.70 &         31.38 & may11 & B &   \\
GB6J015451+362746 &     118.10 &     247.20 &          0.81 & may11 & B &   \\
GB6J021625+400101 &      35.30 &      46.50 &          3.00 & may11 & B &   \\
GB6J022526+371029 &     175.00 &     197.60 &          7.49 & may11 & B &   \\
GB6J061641+663024 &     260.70 &     265.30 &         55.90 & may27 & B &   \\
GB6J065010+605001 &      40.90 &      59.60 &          2.18 & may27 & B &   \\
GB6J065422+504223 &     313.30 &     327.50 &         22.06 & may27 & B &   \\
GB6J065648+560258 &      52.26 &      53.26 &         49.48 & FIRST & B &   \\
GB6J070648+592327 &      49.82 &      51.03 &         37.65 & FIRST & B &   \\
GB6J070932+501056 &     107.73 &     109.36 &         64.83 & FIRST & B &   \\
GB6J071044+422053 &     266.98 &     300.35 &          3.70 & FIRST & B &   \\
GB6J071045+473203 &     802.19 &     894.12 &          3.81 & FIRST & B &   \\
GB6J071510+452554 &      71.99 &      73.53 &         44.43 & FIRST & B &   \\
GB6J071547+413926 &      30.28 &     103.12 &          0.38 & FIRST & B &   \\
GB6J072028+370647 &      19.51 &      20.18 &         26 & FIRST & B &   \\
GB6J072151+712036 &     481.90 &     793.80 &          1.55 & may27 & B &   \\
GB6J072403+535121 &      12.42 &      12.86 &         28.23 & FIRST & B &   \\
GB6J072849+570124 &     437.26 &     462.90 &         11.96 & FIRST & B &   \\
GB6J073328+351532 &      45.56 &     102.30 &          0.68 & FIRST & NVSS &   \\
GB6J073502+475011 &     356.64 &     404.60 &          4.18 & FIRST & NVSS &   \\
GB6J073728+594106 &     445.30 &     459.40 &         30.34 & may27 & B &   \\
  GB6J073758+643048 &     448.18 &     462.18 &         32.01 & FIRST & B &   \\
GB6J073933+495449 &     103.77 &     107.31 &         27.62 & FIRST & B &   \\
GB6J074904+451027 &      79.46 &     112.55 &          2.02 & FIRST & B &   \\
GB6J080036+501047 &     113.94 &     118.17 &         26.94 & FIRST & B &   \\
GB6J080624+593059 &      68.10 &      68.10 &       $>$681 & may27 & B &   \\
GB6J080839+495033 &     901.70 &     930.31 &         12.97 & FIRST & B &   \\
GB6J080949+521856 &     184.03 &     188.29 &         37.96 & FIRST & B &   \\
GB6J081622+573858 &     139.60 &     170.80 &          4.47 & may27 & B &   \\
GB6J082154+503136 &      53.11 &      53.94 &         20.54 & FIRST & B &   \\
GB6J082437+405712 &      87.01 &     174.72 &          0.61 & FIRST & B &   \\
GB6J083055+540041 &      17.80 &      19.10 &         12.89 & FIRST & B &   \\
GB6J083140+460800 &     124.48 &     130.74 &         17.55 & FIRST & B &   \\
GB6J083411+580318 &      54.20 &      58.20 &         12.40 & may27 & B &   \\
GB6J083455+553430 &    7998.29 &    8254.60 &         25.13 & FIRST & B &   \\
GB6J083901+401608 &      32.88 &      38.54 &          4.86 & FIRST & B &   \\
GB6J084215+452547 &      65.53 &      67.67 &         12.73 & FIRST & B &   \\
GB6J084241+391100 &     $<$1.4 &      55.70 &       $<$0.03 & u.l.  & NVSS & double \\
GB6J084352+374232 &      61.98 &      66.59 &          4.92 & FIRST & B &   \\
GB6J085005+403610 &     109.87 &     118.29 &         13.05 & FIRST & B &   \\
   GB6J085008+593054 &      11.56 &      13.07 &          2.83 & FIRST & B &   \\
GB6J085317+682824 &      41.20 &     288.60 &          0.16 & may27 & NVSS &   \\
GB6J085416+532731 &      21.34 &      22.07 &          8.56 & FIRST & B &   \\
   GB6J085449+621900 &     278.00 &     287.75 &         28.51 & FIRST & B(m) &   \\
GB6J090536+470603 &      88.10 &      92.81 &         18.70 & FIRST & B &   \\
GB6J090615+463633 &     306.81 &     313.57 &         41.84 & FIRST & B &   \\
GB6J090650+412426 &      56.00 &      58.21 &         24.66 & FIRST & B &   \\
GB6J090757+493558 &      38.07 &      38.45 &         96.77 & FIRST & B &   \\
GB6J092914+501323 &     495.50 &     503.44 &         62.41 & FIRST & B &   \\
  GB6J092932+625637 &      22.34 &      55.50 &          0.67 & FIRST & NVSS &   \\
GB6J093254+673654 &      23.20 &     229.80 &          0.11 & may27 & NVSS &   \\
GB6J094319+361447 &      73.77 &      74.91 &         63.29 & FIRST & B &   \\
GB6J094542+575739 &      69.28 &      78.29 &          7.69 & FIRST & B &   \\
GB6J094557+461907 &      29.87 &      31.62 &         15.57 & FIRST & B &   \\
GB6J094614+581932 &      25.16 &      61.31 &          0.61 & FIRST & B &   \\
GB6J094832+553538 &      26.22 &      28.59 &          9.89 & FIRST & B &   \\
GB6J095227+504837 &     104.85 &     110.36 &          7.47 & FIRST & B &   \\
GB6J095531+690357 &     104.00 &     127.90 &          4.35 & may27 & B &   \\
GB6J095552+694047 &     664.60 &    6859.80 &          0.11 & may27 & B &   \\
GB6J095736+552258 &    2804.17 &    3056.17 &          5.83 & FIRST & B &   \\

\end{tabular}
\end{table*}
\newpage

\addtocounter{table}{-1}
\begin{table*}
% \begin{minipage}{140mm}
\caption{VLA data}
\begin{tabular}{l r r r l l l}
name & core flux & total flux &  R & data (core) & data (total) & comments \\ 

GB6J100055+533158 &      38.49 &     108.80 &          0.23 & FIRST & NVSS &   \\
GB6J100712+502346 &      29.23 &      30.64 &         20.73 & FIRST & B &   \\
GB6J100724+580201 &     160.16 &     167.47 &         21.91 & FIRST & B &   \\
GB6J101028+413230 &     258.74 &     340.26 &          1.97 & FIRST & B &   \\
GB6J101244+423009 &      69.90 &      79.90 &          6.99 & FIRST & B &   \\
GB6J101504+492606 &     385.85 &     398.16 &         26.12 & FIRST & B &   \\
GB6J101859+591126 &      74.30 &      84.70 &          7.14 & may27 & B &   \\
GB6J102310+394759 &     807.88 &    1078.67 &          1.32 & FIRST & B &   \\
GB6J102521+372641 &      17.47 &      64.20 &          0.35 & FIRST & NVSS &   \\
GB6J103118+505350 &      36.72 &      37.97 &         21.59 & FIRST & B &   \\
GB6J103123+744158 &     148.20 &     185.70 &          3.52 & may27 & B &   \\
GB6J103318+422228 &      26.02 &      33.21 &          3.02 & FIRST & B &   \\
GB6J103550+375646 &      52.76 &      54.26 &         14.02 & FIRST & B &   \\
GB6J103653+444832 &      35.25 &      35.82 &         54.87 & FIRST & B &   \\
GB6J103742+571158 &     127.84 &     130.02 &         58.64 & FIRST & B &   \\
GB6J103951+405557 &      29.94 &      29.73 &       $>$299.40 & FIRST & B &   \\
GB6J104630+544953 &      93.02 &      98.51 &         13.56 & FIRST & B &   \\
GB6J104910+373742 &      55.49 &      56.56 &         12.97 & FIRST & B &   \\
GB6J105115+464439 &     284.21 &     318.38 &          8.32 & FIRST & B &   \\
GB6J105203+424203 &      88.75 &      90.14 &         63.85 & FIRST & B &   \\
GB6J105344+493006 &      52.08 &      68.68 &          2.75 & FIRST & B(m) &   \\
GB6J105430+385500 &      72.06 &      73.07 &         30.19 & FIRST & B &   \\
GB6J105730+405631 &      46.21 &      47.27 &         42.51 & FIRST & B &   \\
GB6J105837+562817 &     207.96 &     221.38 &         13.55 & FIRST & B &   \\
   GB6J110242+594132 &     416.15 &     419.19 &         48.37 & FIRST & B &   \\
GB6J110428+381228 &     557.26 &     768.50 &          2.56 & FIRST & NVSS &   \\
GB6J110508+465311 &      55.88 &      56.18 &        186.27 & FIRST & B &   \\
GB6J110552+394649 &      22.44 &      39.93 &          1.28 & FIRST & B &   \\
  GB6J110657+603345 &      28.51 &      32.40 &          6.50 & FIRST & B &   \\
GB6J110939+383046 &       9.85 &       9.62 &        $>$98.50 & FIRST & B &   \\
GB6J111106+522751 &      39.53 &     115.70 &          0.52 & FIRST & NVSS &   \\
GB6J111206+352707 &      47.21 &      49.21 &         23.04 & FIRST & B &   \\
   GB6J111903+602832 &       9.35 &      60.30 &          0.16 & FIRST & nvss &   \\
   GB6J111912+623938 &      37.26 &      39.23 &         17.04 & FIRST & B &   \\
   GB6J111914+600459 &     186.43 &     191.98 &          9.23 & FIRST & B &   \\
GB6J112047+421206 &      24.63 &      25.09 &         47.63 & FIRST & B &   \\
GB6J112157+431459 &      44.66 &      45.75 &         40.97 & FIRST & B &   \\
GB6J112413+513350 &      43.87 &      49.86 &          5.93 & FIRST & B &   \\
GB6J112832+583322 &      30.00 &     575.60 &          0.05 & A & B & double \\
GB6J113626+700931 &     136.40 &     353.80 &          0.60 & may27 & B &   \\
GB6J113629+673707 &      40.10 &      50.80 &          3.30 & may27 & B &   \\
GB6J114047+462207 &      78.85 &      81.57 &         26 & FIRST & B &   \\
GB6J114115+595309 &      92.30 &      92.70 &        227.92 & may27 & B &   \\
GB6J114300+730413 &      39.30 &      42.90 &          9.72 & may27 & B &   \\
   GB6J114312+612214 &      65.88 &      69.40 &         18.72 & FIRST & B &   \\
GB6J114722+350109 &     608.81 &     637.60 &         19.90 & FIRST & NVSS &   \\
GB6J114850+592459 &     446.40 &     477.90 &         14.02 & may27 & B &   \\
GB6J114856+525432 &      95.57 &     100.43 &          7.47 & FIRST & B &   \\
GB6J114959+552832 &     137.80 &     142.53 &         29.13 & FIRST & B &   \\
GB6J115126+585913 &     137.20 &     192.40 &          2.49 & may27 & B &   \\
GB6J115757+552713 &      93.45 &      98.51 &         18.41 & FIRST & B &   \\
GB6J120209+444452 &      46.97 &     106.00 &          0.80 & FIRST & NVSS &   \\
GB6J120304+603130 &     157.10 &     194.50 &          3.94 & may27 & B &   \\
GB6J120328+480316 &      66.83 &      68.53 &         21.63 & FIRST & B &   \\
GB6J120334+451050 &      31.85 &      33.56 &          9 & FIRST & B &   \\
GB6J120922+411938 &     397.21 &     398.39 &        336.62 & FIRST & B &   \\
GB6J121008+355224 &      16.70 &      25.70 &          1.82 & may27 & NVSS &   \\
GB6J121331+504446 &      99.65 &     102.73 &         31.38 & FIRST & B &   \\
GB6J121510+462710 &     161.18 &     278.59 &          0.80 & FIRST & B(m) &   \\
GB6J121541+361924 &       1.60 &      34.80 &          0.05 & may27 & NVSS &   \\
GB6J121558+354313 &      10.63 &      48.08 &          0.25 & FIRST & B &   \\
GB6J121736+515502 &      78.51 &      80.67 &         17.39 & FIRST & B &   \\
GB6J122208+581427 &      41.20 &      45.60 &          8.51 & may27 & B &   \\
GB6J122306+582659 &      18.90 &     122.50 &          0.18 & may27 & NVSS &   \\
GB6J122405+500130 &      44.44 &      46.22 &         12.11 & FIRST & B &   \\

\end{tabular}
\end{table*}
\newpage

\addtocounter{table}{-1}
\begin{table*}
% \begin{minipage}{140mm}
\caption{VLA data}
\begin{tabular}{l r r r l l l}
name & core flux & total flux & R & data (core)& data (total) & comments \\ 
GB6J123012+470031 &      85.24 &      87.47 &         36.78 & FIRST & B &   \\
GB6J123151+353929 &      40.25 &      40.96 &         49.90 & FIRST & B &   \\
GB6J123350+502630 &      67.42 &     283.30 &          0.26 & FIRST & NVSS &   \\
GB6J123413+475408 &     352.13 &     362.18 &         25.48 & FIRST & B &   \\
GB6J123417+505441 &      50.70 &      52.40 &         25.44 & FIRST & B &   \\
GB6J123532+522839 &      82.14 &      83.92 &         17.42 & FIRST & B &   \\
GB6J124313+362755 &     114.70 &     147.30 &          3.52 & may27 & B &   \\
GB6J124732+672322 &     252.00 &     360.00 &          2.10 & may27 & NVSS &   \\
GB6J124818+582029 &     179.90 &     184.10 &         42.83 & may27 & B &   \\
GB6J125311+530113 &     378.29 &     420.34 &          9 & FIRST & B &   \\
GB6J125614+565220 &     235.28 &     309.90 &          3.02 & FIRST & NVSS &   \\
GB6J130132+463357 &      87.11 &      90.69 &         20.17 & FIRST & B &   \\
GB6J130146+441612 &      51.53 &      53.64 &         24.42 & FIRST & B &   \\
GB6J130836+434405 &      51.85 &      52.27 &        119.32 & FIRST & B &   \\
GB6J130924+430502 &      55.32 &      58.19 &         19.28 & FIRST & B &   \\
GB6J131215+445023 &      28.31 &     139.90 &          0.24 & FIRST & NVSS &   \\
GB6J131218+351522 &      43.92 &      44.72 &         46.37 & FIRST & B &   \\
GB6J131328+363538 &       9.90 &     122.80 &          0.09 & may27 & NVSS &   \\
GB6J131739+411538 &     247.39 &     266.50 &         12.14 & FIRST & NVSS &   \\
GB6J131947+514759 &     345.72 &    1093.30 &          0.22 & FIRST & NVSS &   \\
GB6J132513+395610 &      55.18 &      56.32 &         45.02 & FIRST & B &   \\
GB6J134035+444801 &      81.37 &      82.20 &         92.00 & FIRST & B &   \\
GB6J134139+371653 &      62.89 &      90.38 &          1.96 & FIRST & B &   \\
GB6J134444+555322 &     120.05 &     132.02 &          9.66 & FIRST & B &   \\
GB6J134856+395904 &      79.36 &      80.33 &         81.13 & FIRST & B &   \\
GB6J134913+601114 &      17.50 &      79.30 &          0.28 & may27 & NVSS &   \\
GB6J134934+534125 &     960.41 &    1048.37 &          5.52 & FIRST & B &   \\
GB6J135313+350912 &      41.10 &      42.00 &         40.10 & may27 & B &   \\
GB6J135327+401700 &      36.19 &      38.30 &         17.02 & FIRST & B &   \\
GB6J135607+413637 &      23.47 &      24.31 &         16.46 & FIRST & B &   \\
GB6J141132+742404 &     113.20 &     128.20 &          7.55 & may27 & B &   \\
GB6J141159+423952 &      71.17 &      72.31 &         33.07 & FIRST & B &   \\
GB6J141343+433959 &      42.81 &      43.86 &         37.44 & FIRST & B &   \\
GB6J141536+483102 &      42.79 &      43.61 &         52.18 & FIRST & B &   \\
GB6J141946+542328 &     573.09 &     583.67 &         47.06 & FIRST & B &   \\
GB6J142312+505543 &     117.99 &     139.13 &          4.38 & FIRST & B &   \\
GB6J142814+391222 &     162.92 &     239.45 &          1.69 & FIRST & B &   \\
GB6J143120+395245 &     207.87 &     210.04 &         43.30 & FIRST & B &   \\
GB6J143239+361823 &     110.20 &     116.50 &         17.26 & may27 & B &   \\
   GB6J143646+633645 &     855.43 &     877.84 &         12.44 & FIRST & B &   \\
GB6J143920+371148 &      54.22 &      56.71 &         10.94 & FIRST & B &   \\
GB6J144920+422103 &     159.08 &     165.64 &         20.58 & FIRST & B &   \\
  GB6J150522+604007 &       2.23 &      56.30 &          0.03 & FIRST & nvss &   \\ 
GB6J150854+373318 &      13.18 &      14.71 &          8.36 & may27 & B &   \\
GB6J150914+700436 &      23.80 &      79.80 &          0.42 & may27 & B &   \\
GB6J151806+424346 &       7.0  &      50.80 &          0.16 & A & NVSS & double \\
GB6J151807+665746 &      42.60 &      42.60 &        $>$403.03 & may27 & B &   \\
GB6J151838+404532 &      44.17 &      45.41 &         33.45 & FIRST & B &   \\
GB6J153900+353053 &      87.86 &      89.15 &         63.06 & FIRST & B &   \\
GB6J153931+381011 &      25.52 &      51.60 &          0.92 & FIRST & NVSS &   \\
GB6J154255+612950 &     126.30 &     130.00 &         34.14 & may27 & B &   \\
GB6J154255+705000 &      45.30 &      45.30 &        $>$441.95 & may27 & B &   \\
GB6J154504+525930 &     158.29 &     208.10 &          2.46 & FIRST & NVSS &   \\
GB6J155158+580642 &     192.29 &     200.91 &         10.37 & FIRST & B &   \\
GB6J155848+562524 &     180.61 &     200.31 &          7.05 & FIRST & B &   \\
GB6J155901+592437 &     161.10 &     215.90 &          2.77 & may27 & NVSS &   \\
GB6J160357+573101 &     332.92 &     342.12 &          9.40 & FIRST & B &   \\
GB6J160532+531250 &      26.24 &      62.70 &          0.68 & FIRST & NVSS &   \\
   GB6J160820+601834 &      55.66 &      56.40 &         63.75 & FIRST & B &   \\
GB6J161447+374554 &      49.26 &      51.11 &         10.55 & FIRST & B &   \\
GB6J161941+525617 &     179.07 &     184.76 &          9.42 & FIRST & B &   \\
GB6J161947+523319 &    $<$1.4  &      48.90 &       $<$0.03 & u.l.  & NVSS & double \\
GB6J162308+390946 &     189.57 &     192.46 &         22.09 & FIRST & B &   \\
GB6J162509+405345 &     107.33 &     129.97 &          4.60 & FIRST & B &   \\

\end{tabular}
\end{table*}
\newpage

\addtocounter{table}{-1}
\begin{table*}
% \begin{minipage}{140mm}
\caption{VLA data}
\begin{tabular}{l r r r l l l}
name & core flux & total flux & R & data (core) & data (total) & comments \\ 
GB6J162612+512044 &      20.08 &      50.30 &          0.56 & FIRST & NVSS &   \\
GB6J162636+580914 &     134.84 &     533.70 &          0.19 & FIRST & NVSS &   \\
GB6J163801+552547 &      24.40 &     175.00 &          0.16 & may27 & B(m) &   \\
GB6J163813+572029 &    1005.87 &    1091.69 &          6.69 & FIRST & B &   \\
GB6J164258+394842 &    6050.06 &    6598.61 &          6.92 & FIRST & B &   \\
GB6J164420+454644 &     104.53 &     183.50 &          1.08 & FIRST & NVSS &   \\
GB6J164734+494954 &     103.62 &     178.00 &          1.33 & FIRST & NVSS &   \\
GB6J165138+400227 &      42.73 &      43.90 &         11.01 & FIRST & B &   \\
GB6J165353+394541 &    1394.38 &    1420.36 &         51.92 & FIRST & B &   \\
GB6J165414+434319 &      30.98 &      50.42 &          1.59 & FIRST & B &   \\
GB6J165503+540754 &      16.01 &      49.60 &          0.43 & FIRST & NVSS &   \\
GB6J165547+444735 &      32.57 &      33.26 &         43.87 & FIRST & B &   \\
GB6J165721+570556 &     813.53 &     972.15 &          2.25 & FIRST & B &   \\
GB6J170123+395432 &     250.98 &     254.93 &         63.54 & FIRST & B &   \\
GB6J170449+713840 &      38.40 &      40.20 &         21.33 & may27 & B &   \\
GB6J170716+453607 &     681.65 &     795.00 &          3.65 & FIRST & NVSS &   \\
GB6J171523+572434 &      43.70 &      44.40 &         60.79 & may27 & B &   \\
GB6J171718+422711 &     128.29 &     132.47 &         25.93 & FIRST & B &   \\
GB6J171813+422759 &      83.05 &     111.60 &          2.47 & FIRST & NVSS &   \\
GB6J171914+485839 &     145.66 &     158.66 &         10.94 & FIRST & B &   \\
GB6J171937+480404 &      61.37 &      62.45 &         27.26 & FIRST & B &   \\
GB6J171941+354700 &       7.46 &      34.40 &          0.22 & FIRST & NVSS &   \\
GB6J172110+354217 &     386.51 &     821.00 &          0.70 & FIRST & NVSS &   \\
GB6J172535+585127 &      52.00 &      72.40 &          2.55 & may27 & B &   \\
GB6J172722+551059 &     141.79 &     149.13 &         15.49 & FIRST & B &   \\
GB6J172818+501315 &     210.70 &     220.13 &         21.17 & FIRST & B &   \\
GB6J172859+383819 &     240.27 &     245.41 &         19.56 & FIRST & B &   \\
GB6J173047+371451 &      62.01 &     102.10 &          1.55 & FIRST & NVSS &   \\
GB6J173410+421933 &        8.0  &      50.80 &         0.19 & A  & NVSS & double \\
GB6J174231+594513 &     105.50 &     110.50 &         21.10 & may27 & B &   \\
GB6J174455+554220 &     446.70 &     644.50 &          2.19 & may27 & B &   \\
GB6J174900+432151 &     235.00 &     254.50 &         12.05 & may27 & B &   \\
GB6J175546+623652 &     272.10 &     278.70 &         40.13 & may27 & B &   \\
GB6J175704+535153 &      67.20 &      67.50 &        200.18 & may27 & B &   \\
GB6J175728+552309 &      74.60 &      77.70 &         22.59 & may27 & B &   \\
GB6J175833+663801 &     131.70 &     815.30 &          0.19 & may27 & B &   \\
GB6J180228+481932 &      17.00 &      17.00 &     $>$150.58 & may27 & B &   \\
GB6J180651+694931 &    1192.00 &    1887.10 &          1.63 & may27 & NVSS &   \\
GB6J183850+480237 &      50.50 &      51.70 &         42.08 & may27 & B &   \\
GB6J184033+621257 &      44.90 &      48.50 &         11.88 & may27 & B &   \\
GB6J184941+642522 &      13.0  &     111.00 &         0.13  & A & B & double \\
GB6J185852+682747 &      43.90 &     213.50 &          0.26 & may27 & B &   \\
GB6J191212+660826 &      31.30 &      35.70 &          6.61 & may27 & B &   \\
GB6J194553+705545 &     969.70 &     976.80 &        124.16 & may27 & B &   \\

\end{tabular}

\begin{flushleft}
{\bf Column 2}: the core flux density; 

{\bf Column 3}: the total flux density ;

{\bf Column 4}: the core-dominance parameter R (when the core flux is equal to
the total flux or slightly larger due to the statistical fluctuation, the
computed R is indicated as a lower limit);

{\bf Column 5}: a flag explaining the data used to compute the peak flux density: 
``FIRST'' means from the FIRST data, ``may27'' or ``may11'' are the dates of  
our own observations, ``A'' means from VLA A-array data and ``u.l.'' means
that the core has not been detected and we have assumed
an upper-limit of 1.4 mJy;

{\bf Column 6}: a flag explaining the data used to compute the total
flux density: B means VLA B-array data and 
NVSS  means  from the NVSS data;

{\bf Column 7}: a comment indicating whether the source has a double
morphology;

\end{flushleft}

\end{table*}

\end{document}